\documentclass[12pt]{article}
\usepackage{graphicx}
\usepackage{epsfig}
\usepackage{amsmath}
\usepackage{amssymb}
\usepackage{amsthm}
\usepackage{graphics}

\newcommand{\wt}{\widetilde}
\newcommand{\wh}{\widehat}
\newcommand{\lp}{\left}
\newcommand{\rp}{\right}
\newcommand{\la}{\langle}
\newcommand{\ra}{\rangle}

\newcommand{\tmn}{\la T_{\mu\nu}\ra}
\newcommand{\tud}{\la T_\mu^{~\nu}\ra}
\newcommand{\mpl}{M_{P}}

\newcommand{\af}{$AdS_5$}
\newcommand{\afr}{$AdS_4$}
\newcommand{\at}{$AdS_3$}
\newcommand{\p}{p}

\input epsf%
\begin{document}

\centerline{\Large \bf Strongly Coupled Radiation from Moving Mirrors }%
\vspace{.6cm} %
\centerline{\Large \bf  and Holography in the Karch-Randall model}

\vspace{1cm}

\centerline{\large Oriol
Pujol{\`a}s\footnote{pujolas@ccpp.nyu.edu}}

\vspace{5mm}

\centerline{\emph{Center for Cosmology and Particle Physics}}
\centerline{\emph{Department of Physics, New York University} }
\centerline{\emph{New York, NY 10003, USA} }

\vspace{1cm}

\begin{abstract}

Motivated by the puzzles in understanding how Black Holes
evaporate into a strongly coupled Conformal Field Theory, we study
particle creation by an accelerating mirror. We model the mirror
as a gravitating Domain Wall and consider a CFT coupled to it
through gravity, in asymptotically Anti de Sitter space.
This problem (backreaction included) can be solved exactly at one
loop. At strong coupling, this is dual to a Domain Wall localized
on the brane in the Karch-Randall model, which can be fully solved
as well.
Hence, in this case one can see how the particle production is
affected by A) strong coupling and B) its own backreaction.
We find that A) the amount of CFT radiation at strong coupling is
not suppressed relative to the weak coupling result; and
B) once the boundary conditions in the $AdS_5$ bulk are
appropriately mapped to the conditions for the CFT on the boundary
of $AdS_4$, the Karch-Randall model and the CFT side agree to
leading order in the backreaction. This agreement holds even for a
new class of self-consistent solutions (the `Bootstrap' Domain
Wall spacetimes) that have no classical limit.
This provides a quite precise check of the holographic
interpretation of the Karch-Randall model. We also comment on the
massive gravity interpretation.

As a byproduct, we show that relativistic Cosmic Strings (pure
tension codimension 2 branes) in Anti de Sitter are repulsive and
generate long-range tidal forces even at classical level. This is
the phenomenon dual to particle production by Domain Walls.

\end{abstract}

\newpage

\tableofcontents

\hspace{1cm}

\section{Introduction}

The AdS/CFT correspondence \cite{malda,witten,klebanov} provides a
powerful method to investigate the dynamics of strongly coupled
gauge theories. The correspondence relates a supergravity theory
around 5 dimensional Anti de Sitter space (\af) with a Conformal
Field Theory (CFT) defined on the boundary of \af\ and decoupled
from gravity -- in the limit of large number of colours $N$ with
the 't Hooft coupling $\lambda\equiv g^2_{YM}N\gg1$ fixed.

Dynamical gravity can be included in the 4D theory by breaking
explicitly conformal invariance in the UV, which corresponds to
the introduction of a brane in the \af~bulk
\cite{verlinde,gubser,ahpr}, that is, to the Randall Sundrum (RS)
model \cite{rs2}.
This leads to a `cutoff' version of the correspondence that can be
sharply stated as relating the (classical) 5D solutions of the RS
model with matter localized on the brane to 4D geometries where
the quantum effects from the CFT and their backreaction are taken
into account \cite{efk,takahiroBH}. If true, this extended version
of the correspondence represents an almost tailor-made tool to
study semiclassical gravity problems with strongly
self-interacting fields.

In particular, this has led to a number of interesting claims
concerning how Black Hole (BH) evaporation is affected by the
strongly coupled nature of the field theory.
Ignoring the self-interactions (that is, for $\lambda=0$), one
expects the Hawking radiation to scale as $N^2$. Since the
backreaction from this flux of energy is automatically accounted
for in the braneworld construction, one would conclude that there
should be no static (large) BHs localized on a RS brane
\cite{takahiroBH,efk}. However, as argued in \cite{frw} the
strongly coupled nature of the CFT might render this conclusion
invalid.
If the CFT is in a confining phase, then the BH should actually
emit colour singlet states, implying that there should be no $N^2$
enhancement in the radiation. Given that the 5D gravity dual gives
the leading order contribution in the $1/N$ expansion, this would
lead to zero Hawking radiation at this order.
This is supported \cite{frw} by the (warped) uniform Black String
(BS) solution \cite{chr} in the two-brane RS model \cite{rs1}.
This solution is stable for a large enough horizon radius, and
displays no Hawking radiation from the 4D point of view. However,
the dual of the two-brane RS model is a `CFT' where conformal
invariance is broken in the IR even in flat space (one is not
really dealing with a massless theory), so it still seems unclear
what happens for an unbroken CFT, that is, in the one-brane RS
model (see also
\cite{balbinotFabbri,FabbriProcopio,fabbriolmo2,takahiroFloating,tamaInitialData}).

A sharper situation arises for asymptotically $AdS_4$ Black Holes.
The presence of a negative cosmological constant $\Lambda_4$
effectively places the system in a box,  and the CFT can reach an
equilibrium configuration with the BH. In this way one can avoid
the technical problems associated with a putative time dependence
of the asymptotically flat case. The correspondence in the
presence of $\Lambda_4<0$ is perhaps not as well understood as for
$\Lambda_4=0$, but the picture is that 4D gravity with
$\Lambda_4<0$ and the strongly coupled CFT is dual to the
Karch-Randall (KR) model \cite{kr} with either one or two branes,
depending on the boundary conditions for the CFT fields at the
$AdS_4$ boundary.
The one-brane model arises with the so-called Karch-Randall
boundary conditions (see below)
\cite{massGaugeInv,boussoRandall,massimoHiggs,duffLiuSati,massimoReview,DeWolfe2}.
Instead, with reflecting boundary conditions, one obtains a
two-brane model with an additional $Z_2$ symmetry across the bulk.
As shown in \cite{grz} (see also \cite{gregory}), with reflecting
boundary conditions the vacuum polarization at strong coupling
dramatically differs from the weak coupling result. Indeed, for
large enough horizon radii the uniform Black String solution is
stable \cite{hirayama,ChamblinKarch}, so this should be the
physical solution. Clearly, this solution corresponds to a state
of the CFT where the thermal component of $\tmn^{CFT}$ vanishes.
Since the one loop estimate is instead of order $N^2$ \cite{grz},
one concludes that the `shutdown' of the Hawking radiation must be
a strong coupling effect.
It is worth pointing out that even though the
Schwarzschild-$AdS_4$ Black String is stable for the one-brane KR
model as well, this solution does not seem relevant because the
bulk is asymptotically substantially different from $AdS_5$.
Hence, with KR boundary conditions ({\em i.e.}, again, in the
one-brane case)
we do not expect a similar suppression of the radiation.\\

The purpose of the present article is to gain some insight by
exploring the particle creation phenomena that occur in a toy
model consisting of a Domain Wall (DW) in $AdS_4$ with a CFT
probing it (through gravity). The reason to choose this kind of
source is that, once its gravitational effect is accounted for, a
DW \emph{is} a physical implementation of an accelerating mirror,
hence one expects analogous particle production generically.
The great advantage of this case is that this problem can be
solved exactly both at 1-loop and at strong coupling, where it
reduces to finding the 5D solution for a DW localized on the brane
in the Karch-Randall model.
Furthermore, in both cases it is possible to separate the problem
in two steps, by first neglecting the backreaction and then
including it.\footnote{The gravity dual for the strongly coupled
case without the backreaction, according to the `standard' AdS/CFT
correspondence, reduces to finding the 5D geometry whose boundary
is conformal to the DW background.}
As a result, one can give a clear account of how the particle
production is affected by A) the 't Hooft coupling (that is,
whether the CFT is weakly or strongly coupled) and B) the
backreaction of the CFT quantum effects themselves.
%


As we shall see, the most interesting case is when the DW is in
asymptotically $AdS_4$ and its tension $\sigma$ is small enough so
that its proper acceleration is less than the $AdS_4$
curvature scale. 
We shall refer to these as \emph{subcritical} DWs.
Finding the amount of produced radiation and in fact the whole
$\tmn^{CFT}$ in the DW background can be done along the lines of
\cite{gripu}, and for subcritical walls reduces to an `ordinary'
Casimir energy computation. We perform this explicitly at one loop
for the two types of relevant boundary conditions (see Section
\ref{sec:quantum}). As a result one obtains a certain amount of
zero temperature CFT radiation in equilibrium with the $AdS_4$
boundary and the mirror (the DW), with an energy density that
`piles up' at a characteristic distance from the DW (see Fig.
\ref{fig:picture}).

We then compare this to the non-perturbative strong coupling
computation that one can infer from the 5D dual (Section
\ref{sec:noBrane}). Our first main result is that for every given
boundary condition, the amount of radiation at weak and at strong
coupling are of the same order, in some cases matching to within a
few per cent (see Fig. \ref{fig:T_p_ratios}). Hence, strong
coupling effects do not substantially suppress the amount
of radiation produced by DWs. \\

One important issue in this discussion concerns the boundary
conditions. The Karch-Randall  choice is specified by allowing the
CFT to communicate at infinity with another field theory (the
``CFT$'\,$") with transparent boundary conditions
\cite{massGaugeInv,boussoRandall,massimoHiggs}. In principle, this
leaves a degree of arbitrariness in that we should specify the
state for the CFT$'$. In our case, this becomes manifest because
these boundary conditions are labelled by a continuous parameter
representing whether the CFT$'$ is probing a ``Domain Wall$'\,$"
({\em i.e.}, wether it is defined on a DW background).
Of course, the natural choice is that the CFT$'$ is in the ground
state \cite{massGaugeInv,massimoHiggs} and that there is no DW$'$
(more in general, that the CFT$'$ lives on a space conformal to
\emph{pure} $AdS_4$). With these boundary conditions one always
finds some radiation, roughly proportional to the DW tension
$\sigma$. Strictly speaking, though, there is always a choice of
boundary conditions for which no radiation is present. This
happens if the CFT$'$ is postulated to probe a DW$'$ with tension
opposite to that of `our' DW. Clearly, though, this is not the
most natural condition. In the 5D gravity dual, this corresponds
to a 5D solution that is not asymptotically global $AdS_5$ but
rather $AdS_5$ with a wedge removed.\footnote{This situation has a
direct parallel in the $AdS_4$ BH case with KR boundary
conditions. The Uniform Warped Black String (even if stable) is
not asymptotically $AdS_5$. This would correspond to a state where
the CFT$'$ is directly probing a ``BH$'$" with the same mass as
the on probed by the CFT.}
Thus, it is also clear from the 5D perspective that this is not
the relevant boundary condition and, rather, asymptotically
globally $AdS_5$ (or the CFT$'$ in the $AdS_4$ ground state) is preferred.\\

%

Regarding the backreaction, the most important point is that once
the parameters \emph{and} the boundary conditions in the two sides
of the correspondence are appropriately mapped, the two
descriptions agree to leading order in the backreaction (which
goes along the lines of previous claims
\cite{no1,no,shiromizu1,shiromizu2,gripu}).
This represents a quite solid check of the cutoff AdS/CFT
correspondence and in particular of the holographic interpretation
of the Karch-Randall model, stemming from the fact that it is
based on exact solutions on the two sides. Indeed, all the
previous checks of the 4Dgravity+CFT/braneworld equivalence are
based either on perturbative arguments
\cite{gt,DuffLiu,takahiroBH,balbinotFabbri,massimoHiggs,massGaugeInv,giannakisLiuRen,duffLiuSati},
on the trace anomaly (and hence are insensitive to the actual
vacuum state)
\cite{hhr,no1,no,shiromizu1,shiromizu2,gripu,takahiroFRW,casadio},
or on lower dimensional models
\cite{ehm1,ehm2,anberSorbo1,anberSorbo2}, while in some known
exact solutions in the 5D side \cite{ks} the CFT computation is
not known.

This match is especially remarkable for the `Bootstrap DW
spacetimes', a new kind of self-consistent solutions of the
semiclassical Einstein equations that do not have a classical
limit.
In these solutions, the Casimir energy is nonzero because of the
nontrivial geometry and the geometry is nontrivial because of the
nonzero Casimir energy, in a way that is self-consistent and under
control in the effective theory.
It is perhaps not so surprising but yet quite revealing that these
solutions also exist in the 5D dual, giving additional evidence
that the braneworld models allow for a semiclassical 4D gravity
interpretation.

The inclusion of the backreaction is also relevant for the massive
gravity interpretation. As is well known, with KR boundary
conditions for the CFT, the graviton acquires a mass${}^2$ of
order $N^2 / (\mpl^2\ell_4^4)$ (where $\ell_4$ is the $AdS_4$
curvature radius) while with reflecting conditions the graviton is
massless
\cite{massGaugeInv,massimoHiggs,duffLiuSati,massimoReview}.
Since the graviton mass is ultimately a quantum effect, one
expects that there is a trace of it in the the radiation produced
by the DWs.
The most natural manifestation of a graviton mass in our setup is
in the form of a \emph{screening} phenomenon, by which we mean
that the effective gravitational effect due to the DW tension as
perceived by far away observers may be smaller than expected.
As we will see, the backreaction from $\tmn^{CFT}$ precisely acts
so as to screen the DW tension in this sense. This effect is not
completely distinctive, though, since a Casimir energy is expected
to be present generically for any choice of boundary conditions.
However, it is possible to compare how much screening it leads to
for each boundary conditions. Interestingly enough, we will find
that for KR conditions there is always more such screening.\\

Finally, we shall comment on a more general point which gives a
quite valuable insight, namely on the gravitational field of a
Cosmic String (a relativistic codimension 2 brane) in $AdS$. This
is relevant to our problem because in the 5D gravity side the DW
localized on the brane is one such object, that is attached to a
codimension 1 brane. Many of the properties of our 5D solutions
simply follow from the peculiarities of isolated Cosmic String
(CS) in $AdS$, so it proves very illustrative to momentarily
dispose of the codimension 1 brane.

One might have expected that the solution representing a CS in
$AdS$ is simply given by $AdS$ with a wedge removed, a space
isometric to $AdS$ that only differs in its global structure and
where  as usual all the gravitational effects would arise through
the deficit angle only. It turns out, though, that the situation
is much richer because, in contrast with the flat space case,  in
asymptotically $AdS$ there is a number of possible boundary
conditions. The locally $AdS$ solution is asymptotically $AdS$
minus a wedge. But for the same given CS tension, there is another
solution, which we shall explicitly construct, that approaches
asymptotically \emph{global} $AdS$. There is of course a continuum
of solutions interpolating between the two, but the asymptotically
global $AdS$ is clearly special, since the gravitational effect
from the CS is localized only in this case. Aside from a conical
singularity, these solutions display a non-zero Weyl curvature,
implying that with these boundary conditions, \emph{Cosmic Strings
produce tidal forces}. Not only that, they also generate a
(repulsive) Newtonian potential. This is in fact the dual
phenomenon lying behind the particle production by DWs in 4D.

The reason why there is more than one possible boundary condition
in asymptotically $AdS$ space is linked to the presence of a very
interesting set of everywhere-regular vacuum solutions with
$\Lambda<0$, the so-called Hyperbolic $AdS$ Solitons. These
solutions approach asymptotically $AdS$ with a wedge removed/added
and have a nonzero Weyl curvature.
In fact, these solitons can be viewed as (purely gravitational
versions of) Cosmic Strings, since they are lumps of curvature
localized around a codimension 2 region of spacetime and generate
a deficit/excess angle.
The globally $AdS$ CS solutions of the previous paragraph are
simply the superposition of an ordinary (material) CS and one of
the $AdS$ solitons. It is always possible to compensate the CS
tension with the effective `tension' carried by the gravitational
soliton, in such a way that there is no deficit angle at
infinity, where instead only the Weyl curvature is left.\\
%
%

This paper is organized as follows. Section \ref{sec:cft} is
devoted to the four dimensional side of the correspondence, having
in mind a weakly coupled CFT. After a brief review of the
connection between Domain Walls and accelerating mirrors, we
discuss the quantum effects in DW backgrounds (that is ignoring
the backreaction) in Section \ref{sec:quantum}.
In Section \ref{sec:bc} we motivate and describe the two kinds of
boundary conditions that we shall consider. In Section
\ref{sec:ECas_weak} we perform the one loop computation of the
Casimir energy on $AdS_3\times S_1$, which is the background
relevant for the subcritical walls. We include the backreaction in
Section \ref{sec:Backreact}. In Section \ref{sec:massgrav}, we
discuss the massive gravity interpretation and in in Section
\ref{sec:qs} we describe the Bootstrap DW spacetimes. Section
\ref{sec:5D} deals with the strong coupled CFT, first in Section
\ref{sec:noBrane} without dynamical gravity, that is with no
branes. We infer the strong coupling version of the Casimir energy
in Section \ref{sec:ECas_strong}. The dual with dynamical gravity,
that is, the DW localized on the brane in the KR model is dealt
with in Section \ref{sec:DWs in KR}. Finally, in Section
\ref{sec:CS} we discuss the gravitational effects of isolated
codimension 2 branes in asymptotically $AdS$ space, and
we conclude in Section \ref{sec:concl}.

\section{CFT radiation from Domain Walls}
\label{sec:cft}

In this Section we discuss particle creation by Domain Walls in
asymptotically $AdS_4$ assuming that the produced quanta belong to
a weakly coupled four dimensional CFT. We perform our analysis in
two steps: in Section \ref{sec:quantum} we compute the amount of
produced radiation ignoring its backreaction on the geometry,
which we take into account in Section \ref{sec:Backreact}.

\subsection*{Preamble: Domain Walls as Accelerating Mirrors}%
\addcontentsline{toc}{subsection}{${}$~~~~~~~Preamble: Domain Walls as Accelerating Mirrors}%
\label{sec:mirrors}

\renewcommand{\thesubsection}{\arabic{section}.\Alph{subsection}}

Let us start by briefly reviewing some aspects of the spacetimes
produced by Domain Walls and their similarities with accelerating
mirrors (a more detailed discussion in the same context can be
found in \cite{gripu}).
For definiteness, we assume a negative cosmological constant and
for the moment we ignore the CFT.

In the thin wall approximation, the stress tensor of a
relativistic Domain Wall is
\begin{equation}\label{tmunuThin} %
T^{DW}_{\mu\nu}= \sigma \, \delta(y) \; {\rm
diag}\lp(1,-1,-1,0\rp)_{\mu\nu}
\end{equation}
where $\sigma$ is the tension, and $y$ is the proper coordinate
transverse to the wall. We shall consider the maximally symmetric
configurations, where the metric can be foliated as
\begin{equation}\label{metric}
  ds^2_4=dy^2+R^2(y)ds_{\kappa}^2
\end{equation}
where $ds_{\kappa}^2$ denotes the line element of a 3D Minkowski
($\kappa=0$), de Sitter ($\kappa=1$) or Anti de Sitter
($\kappa=-1$) spacetime of unit radius.

Assuming $Z_2$ symmetry across the DW, the Einstein equations
imply that the extrinsic curvature of the DW $K_{0}\equiv
-(R'/R)|_{0+}$ is
\begin{equation}
\label{K0} %
K_{0} = {\sigma \over 4\mpl^2}
\end{equation}
with $\mpl^2=1/(8\pi G_N)$. This is an \emph{acceleration} scale,
namely the acceleration with which the DW recedes away from
inertial observers. Hence, once gravity is `turned on' a DW
automatically accelerates fuelled by its own surface energy
density, which gives a physical implementation of a moving mirror.
Had we chosen an equation of state on the DW different from
\eqref{tmunuThin}, the acceleration would be time dependent. But
with only the tension term its proper acceleration is constant,
leading to a model of a \emph{uniformly} accelerated mirror.
Needless to say, to actually make of the DW a real mirror that
leads to particle production, one should specify appropriate
boundary conditions for the
propagating quantum fields on the wall. We comment on this below.\\

For the moment, let us mention some other properties of the  DW
spacetimes \eqref{metric}. The equations of motion also determine
the (intrinsic) curvature scale $H^2_{0}\equiv \kappa / R^2_0$ of
the DW as

\begin{equation}\label{GC}
H_{0}^2 = K_{0}^2 -{1\over \ell_4^2}
\end{equation}
where $\ell_4^2 = -3\mpl^2 / \Lambda_4$ and $R_0\equiv R(0)$.
Hence, in the presence of a negative cosmological constant there
is a `critical' value of the DW tension
\begin{equation}\label{sigmac}
\sigma_c \equiv {4 \mpl^2\over \ell_4}~,
\end{equation}
corresponding to an acceleration equal to $1/\ell_4$.
For $|\sigma|$ smaller, equal or larger than $\sigma_c$, the DW
worldvolume is \at, flat or $dS_3$ respectively. In each case, the
`warp' factor is
\begin{equation}\label{Rcases}
R(y)=\left\{%
\begin{array}{llll}
    \ell_4 \, \cosh\lp[ (y_{-} - |y|)/\ell_4 \rp] & \hbox{for} &\kappa=-1 & \hbox{(subcritical)}, \\
    \ell_4 \, e^{ - |y|/\ell_4}, & \hbox{for} &\kappa=0 & \hbox{(critical)},\\
    \ell_4 \, \sinh\lp[ (y_{+} - |y|)/\ell_4 \rp], & \hbox{for} &\kappa=1 & \hbox{(supercritical)}. \\
\end{array}%
\right.
\end{equation}
The integration constants $y_\pm$ depend on $\sigma$, but they are
irrelevant for the present discussion.

The point that we wish to stress here is that the spacetime given
by \eqref{Rcases} contains acceleration horizons for supercritical
and critical cases only (at $y\to\infty$ in the latter case). For
subcritical walls, instead, the horizon is replaced by a bounce in
the warp factor which takes place at finite distance from the
wall.

\begin{figure}[t]
\begin{center}
  \includegraphics[height=4cm]{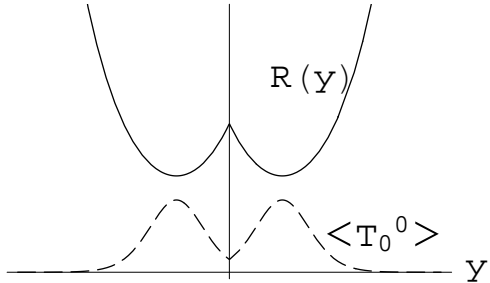}%
\end{center}
\caption{ Schematic picture of a subcritical Domain Wall in
$AdS_4$, effectively acting as a uniformly accelerated mirror with
acceleration smaller than $1/\ell_4$. The warp factor $R(y)$ (see
case $\kappa=-1$ in \eqref{Rcases}) bounces and grows
exponentially. The expectation value of the stress tensor $\tmn$
for the CFT does not vanish. In the equilibrium configuration, the
energy density in the radiation peaks around the bounce of
$R(y)$.}
\label{fig:picture} %
\end{figure}

In all cases the wall is accelerated, so one always expects some
sort of particle production.
Certainly, only when the acceleration $K_0$ exceeds $1/\ell_4$ the
radiation from the wall may be \emph{thermal}, with a temperature
given by $H_0/2\pi = \sqrt{K_0^2-1/\ell_4^2}/2\pi$
\cite{npt,deserLevin} (see \cite{russoTownsend} for a recent
discussion).
%
%
%
However, this does not mean that for subcritical walls there is no
radiation but rather that it is simply not thermal. (We shall be a
bit more precise about this in Section \ref{sec:quantum}) This
does not exclude, for instance, that there is a Casimir energy,
which is what we shall argue that happens in this case. The
resulting situation is depicted in Fig. \ref{fig:picture}: for
subcritical walls, the vacuum expectation value of the CFT energy
density does not vanish and peaks at the location of the bounce.
Arguably, one can view this as an equilibrium configuration where
the produced radiation is in equilibrium with the wall and the
$AdS$ boundary.

Let us add finally that in the usual treatment of particle
creation by moving mirrors, some sort of coupling between the
quantum fields and the mirror is assumed, which can be encoded in
the boundary condition for the fields at the location of the
mirror. Here, we will assume that the CFT fields do not actually
couple directly to the wall, so that on the DW we have transparent
boundary conditions. As we shall see, even with this choice there
is a non-trivial Casimir effect, because the global structure of
the spacetime with the DW is slightly different than without it.
Hence, this represents only a minimal choice, and the addition of
any explicit coupling to the DW is not expected to change the
picture qualitatively.

\subsection{Quantum conformal fields on DW backgrounds}
\label{sec:quantum}

In this Section we compute the expectation value (vev) of the
stress tensor $\tmn$ for conformally coupled fields of any spin in
the Domain Wall spacetimes given by \eqref{Rcases}. For the
moment, we take these as fixed backgrounds -- the backreaction
from $\tmn$ on the geometry is deferred to Section
\ref{sec:Backreact}.

The vev of the stress tensor can always be split  as
\begin{equation}\label{tmncft}
  \tmn^{CFT}= \tmn^{(0)}+ T_{\mu\nu}^{\mathcal{A}}
\end{equation}
where the state-dependent part $\tmn^{(0)}$ encodes the particle
creation or vacuum polarization effects and is tracefree for
conformal fields while  $T_{\mu\nu}^{\mathcal{A}}$ is the
state-independent anomalous contribution. On the spacetime
\eqref{metric}, this only acts so as to renormalize the
cosmological constant and the DW tension \cite{gripu} and will be
considered in more detail in Sec \ref{sec:Backreact}.

Further assuming that the vacuum state of the CFT respects the
symmetries of the background \eqref{metric} implies that
$\tmn^{(0)}$ is of the form
\begin{equation}
    \tud^{(0)}=P(y)\, {\rm diag} \big(\frac{1}{3},-\frac{1}{3},-\frac{1}{3},1\big)_\mu^{~\nu}\,,
    \label{statedep.T}
\end{equation}
where $P$ depends on the direction transverse to the DW only. Now,
the local conservation of the stress-energy tensor demands that
\begin{equation}\label{P}
P(y)={P_0\over R^4(y)}~,
\end{equation}
for some constant $P_0$, which contains all the non-trivial
particle production effects in this problem $\tmn^{CFT}$.
The value of $P_0$ depends on the kind of DW ({\em i.e.}, on
$\kappa$) and on the boundary conditions imposed on the CFT.

From \eqref{P}, it is clear that when the spacetime contains a
horizon, that is for critical and supercritical walls, there is no
radiation of CFT quanta (that is, $P_0=0$).
Otherwise, $\tmn^{CFT}$ would diverge at the horizon, where
$R(y)=0$ \cite{gripu}.\footnote{This is confirmed by explicit
computations at one-loop level \cite{cd,frolovSerebriany,cw1,pt}.}
However, for subcritical walls $R(y)$ is nowhere zero and $P_0$
can be non-trivial. In the next Subsection we compute it at
one-loop as a function of the DW tension, or equivalently its
acceleration.

Note that in principle there can be self-consistent solutions
({\em i.e.} with the backreaction fully taken into account) for
critical or supercritical walls where the horizon is replaced by a
bounce as well. For these, $P_0$ can be nonzero -- in fact it must
be so because it is precisely the Casimir energy that supports the
bounce. Given the key role played by the quantum effects in this
kind of solutions, we shall call them `Bootstrap' solutions. We
discuss them further in Section \ref{sec:qs}. At this point let us
just mention that these solutions are perfectly trustable as long
as $P_0$ is large enough and $\Lambda_4<0$.

Let us add that the distinction whether the stress tensor
\eqref{statedep.T} should be viewed as describing the quanta
created by the wall/mirror or simply as a Casimir energy is a bit
obscure.
Certainly, \eqref{statedep.T} appears quite different from the
radiation perfect fluids commonly considered in cosmology -- it is
not isotropic and the equations of state are $p/\rho=-1$ and $3$
in the longitudinal and transverse directions respectively.
This suggests that \eqref{statedep.T} does not describe a
\emph{thermal} bath of radiation. However, when there is a horizon
this is not the whole story.
Let us accept momentarily for the sake of the argument the
singular vacua with $P_0\neq0$ for supercritical DWs
(alternatively, one could consider non-conformal fields, for which
$\tmn^{(0)}$ does not vanish and is regular at the horizon
\cite{pt}). Neglecting any possible backreaction at the horizon,
the stress tensor in the Milne region would be given by the
analytic continuation of \eqref{statedep.T}. Since in this
continuation $y$ becomes the time coordinate, in the Milne region
$\tmn^{(0)}$ becomes precisely an ordinary radiation perfect
fluid, which one expects to describe a thermal distribution of
quanta with temperature $\sqrt{K_0^2-1/\ell_4^2}/2\pi$
\cite{npt,deserLevin,russoTownsend} despite looking non-thermal in
the Rindler region ({\em i.e.} outside the light-cone defined by
the DW).
For subcritical walls there is no continuation to do, which agrees
with the expectation that \eqref{statedep.T} does not describe
thermal radiation. Still, it can be viewed as a kind of radiation
in equilibrium with the wall and the $AdS$ boundary.

\subsubsection*{Sub-critical Walls}

Let us consider in more detail the sub-critical walls (case
$\kappa=-1$ in \eqref{Rcases}), for which non-trivial quantum
effects are expected.
It is convenient to rewrite the metric \eqref{metric} in terms of
the conformal coordinate $\eta=\int_0^y dy'/R(y')$,
\begin{equation}\label{confcoord}
  ds_4^2={R^2(\eta)}\,\lp(d\eta^2+\,ds_{AdS_3}^2\rp)~.
\end{equation}
Given that $R(y)$ does not vanish and it grows exponentially for
large enough $y$, the range of the conformal coordinate is finite,
\begin{equation}\label{Deta}
  \Delta\eta=\pi+2\arcsin\lp({\sigma\ell_4\over4\mpl^2}\rp)~.
\end{equation}
Note that $\Delta\eta$ is conformally invariant (in other words,
DW spacetimes with different tensions are \emph{not} conformal to
each other) and hence is a parameter that the CFT can be sensitive
to. This quantity, then, is very convenient ({\em e.g.}, it is
conformally invariant) to characterize the DW spaces, and will
play a key role in this discussion.

From \eqref{confcoord} and \eqref{Deta}, we see that every DW
space is conformal to $I\times AdS_3$, where $I$ denotes an
interval of length $\Delta\eta$. Hence, the computation of $\tmn$
is equivalent to a Casimir effect between hyperbolic `plates'
separated a distance $\Delta\eta$.

As is well known, the vevs of the stress tensor for conformal
coupled fields in two conformally related spaces $g_{\mu\nu}$ and
$\wt g_{\mu\nu}$ satisfy
\begin{equation}\label{cftf}
\la T_{\mu\nu}\ra={\sqrt{-\wt g} \over\sqrt{-g}} \, %
\Big[\la \wt T_{\mu\nu} \ra- \wt T_{\mu\nu}^{\cal A} \Big]+
T_{\mu\nu}^{\cal A}~.
\end{equation}
Comparing this with \eqref{tmncft} we see that the first term in
the rhs is the state-dependent part $\tmn^{(0)}$.
Hence, what we need to find is the state-dependent part of $\la
\wt T_{\mu\nu} \ra$ on $I\times AdS_3$. This again takes the form
\eqref{statedep.T} with $\wt P(y) = \wt P_0 / R_*^4$ where  $R_*$
is the radius of the $AdS_3$ factor and, from Eq. \eqref{cftf},
$\wt P_0 =P_0$.
In order to explicitly obtain $P_0$, one could proceed with a
direct mode-summation as was done in \cite{norman} for
$AdS_4\times I$. Here, though, we shall take a slightly quicker
route that is allowed for by the kind of boundary conditions that
we shall impose on the field theory.

\subsubsection{Boundary Conditions}
\label{sec:bc}

Needless to say, the precise form of $P_0(\Delta\eta)$ depends on
the boundary conditions on the DW and at infinity, that is, at the
boundary of the interval $I$.
As mentioned above, we shall take the minimalistic assumption that
the CFT does not couple directly to the DW ({\em i.e.}, the
coupling is through geometry that it produces only), which can be
stated as imposing transparent boundary conditions on the wall.

Because the boundary of $AdS$ is timelike, to proceed we also need
to specify the boundary conditions at infinity.
As explained in
\cite{massimoHiggs,massGaugeInv,duffLiuSati,massimoReview}, these
play a key role in the dual of the Karch-Randall (KR) model
\cite{kr}, in particular to obtain the graviton mass.
Let us describe next the two types of boundary conditions that we
shall consider in this paper.

\subsubsection*{Karch-Randall boundary conditions}

In order to compare the 1-loop with strong coupling results of
Section \ref{sec:5D}, we shall impose the boundary conditions that
are built-in in the Karch-Randall (KR) model \cite{kr}. The
so-called KR boundary conditions are such that the CFT
communicates at infinity with an additional conformal field theory
(which we shall refer to as the CFT$'$) living on an adjacent copy
of $AdS_4$, with transparent boundary conditions
\cite{kr,massGaugeInv,massimoHiggs,massimoReview}.

\begin{figure}[tb]
\begin{center}
  \includegraphics[height=8cm]{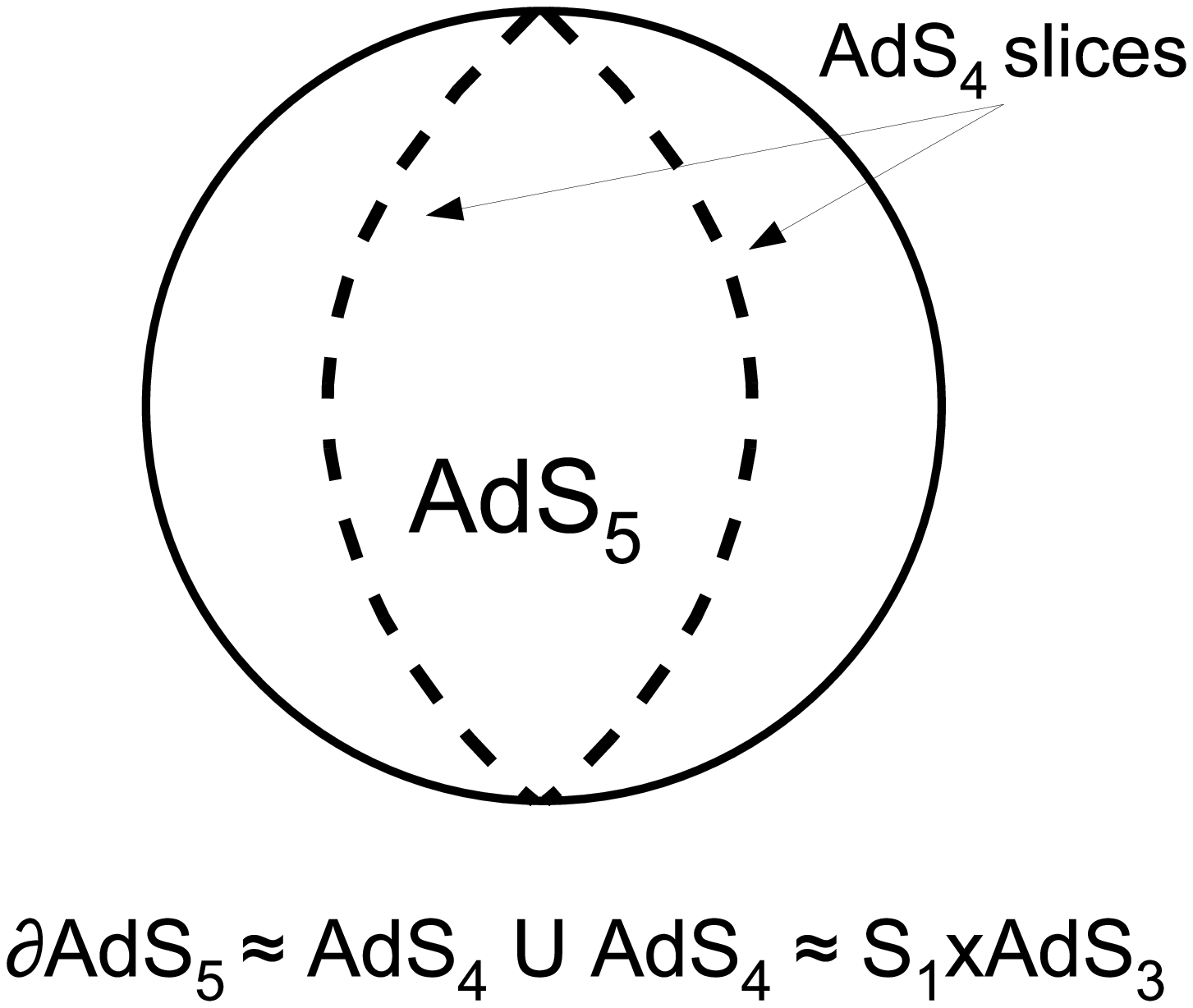}%
\end{center}\vspace{-1cm}
\caption{Five dimensional Anti de Sitter space can be represented
as a disk, with every point representing an \at. The boundary at
infinity has a topology $AdS_3\times S_1 \simeq S_3 \times R$. An
\afr~  slice `covers' only half of the boundary. This motivates
the Karch-Randall boundary conditions.}
\label{fig:KR} %
\end{figure}

In our problem, this means that we have to glue the interval $I$
to another one $I'$ spanned by, say, $\eta'$. Since the boundary
of the two intervals is common, this naturally defines a circle
$S_1 = I \cup I'$ (we will call $\theta$ the coordinate along
$S_1$). `Transparent' boundary conditions at the common boundary
of $I$ and $I'$ then translates into (anti)periodic boundary
conditions on the $S_1$.
Intuitively, this is because the full boundary of $AdS_5$ is
$R\times S_3$, which is equivalent to $AdS_3 \times S_1$, or to
two $AdS_4$ spaces sharing their common boundary, as depicted in
Fig \ref{fig:KR}.

In practice, then, computing the Casimir energy on $AdS_3\times I$
with KR boundary conditions is equivalent to computing it on
$AdS_3 \times S_1$,
\begin{equation}\label{s1ads3}
  R_*^2\lp(d\theta^2+ds_{AdS_3}^2\rp)~,
\end{equation}
(the overall scale $R_*$ is going to be irrelevant)
with the length of the circle given by
\begin{equation}\label{general}
  \Delta\theta=\Delta\eta+\Delta\eta'~.
\end{equation}
In this article, we will assume periodic boundary conditions for
the bosons and antiperiodic for the fermions. In Section
\ref{sec:ECas_strong} we shall comment on the case when the
fermions are periodic.

Note that the transparency of the assumed boundary conditions at
the $AdS_4$ boundary implies that the Casimir energy $P_0$ is
going to be really a function of $\Delta\theta$ and it will depend
on $\Delta\eta$ through Eq. \eqref{general}. Hence, the actual
form of $P_0(\Delta\eta)$ is subject to the choice of boundary
condition encoded by $\Delta\eta'$, which hiddenly entails a
choice of the state for the CFT$'$. Of course, the natural choice
is that the CFT$'$ is in its ground state
\cite{massGaugeInv,massimoHiggs}, which demands in particular that
it is not directly probing any ``Domain Wall$'\,$"  and hence,
\begin{equation}\label{natural}
\Delta\eta'=\pi~.
\end{equation}
With this boundary conditions, then the Casimir energy is going to
be
\begin{equation}\label{P0_KR}
P_0^{KR}(\Delta\eta)\equiv
P_0\Big(\Delta\theta=\Delta\eta+\pi\Big)~
\end{equation} %
(we defer to Section \ref{sec:ECas_weak} the computation of
$P_0(\Delta\theta)$). As we shall see, Eq. \eqref{natural} is the
only choice that leads to no particle production ($P_0=0$) in the
absence of the DW.

By the same token, this also means that there is always a state (a
choice of $\Delta\eta'$) such that there is no particle production
for any DW tension. This will be trivially accomplished choosing
$\Delta\eta'=2\pi-\Delta\eta$ simply because as we will see,
$P_0(\Delta\theta=2\pi)=0$. However, this should not be ascribed
to any strong coupling effect, of course. Rather, in these states
the boundary conditions for the CFT$'$ are quite exotic -- it is
probing the presence of a DW$'$ with tension equal to $-\sigma$.

The choice of vacuum for the CFT$'$ is thus an essential
ingredient of the computation. In the gravity picture, it is dual
to the choice of boundary conditions in the bulk and the natural
condition is going to be that the bulk is asymptotically
\emph{global} $AdS_5$. As we shall see, this corresponds precisely
to the state where the CFT$'$ is in the ground state, Eq.
\eqref{natural}, or more precisely that the geometry that it
probes is conformal to pure $AdS_4$. This goes very much along the
lines of what was observed in \cite{anberSorbo2} for the shock
waves in an \at~ brane, where two solutions with different
asymptotics in the bulk correspond to two rather different states
of the CFT.

\subsubsection*{Reflecting boundary conditions}

Without much additional effort, it is also possible to identify
the form of $\tmn$ for a certain type of reflecting boundary
conditions.
In our setup, reflecting means that on the boundary of $I$ the
fields obey either Dirichlet or Neumann boundary conditions. Given
that $I$ has two boundaries (corresponding to the two halves of
the $R\times S_2$ boundary of $AdS_4$), and that the CFT has a
number of fields of different spin, there are in principle several
different kinds of reflecting boundary conditions. The type that
we will consider is the one that can be obtained as a particular
case of KR conditions by taking the CFT$'$ sector to be identical
to the CFT and letting it probe a replica of `our' DW. In other
words, we simply take
\begin{equation}\label{reflecting}
  \Delta\eta'=\Delta\eta~,
\end{equation}
and so the Casimir energy is
\begin{equation}\label{P0_refl}
P_0^{refl}(\Delta\eta)\equiv
P_0\Big(\Delta\theta=2\Delta\eta\Big)~.
\end{equation} %

\begin{figure}[t]
\begin{center}
  \includegraphics[height=2.5cm]{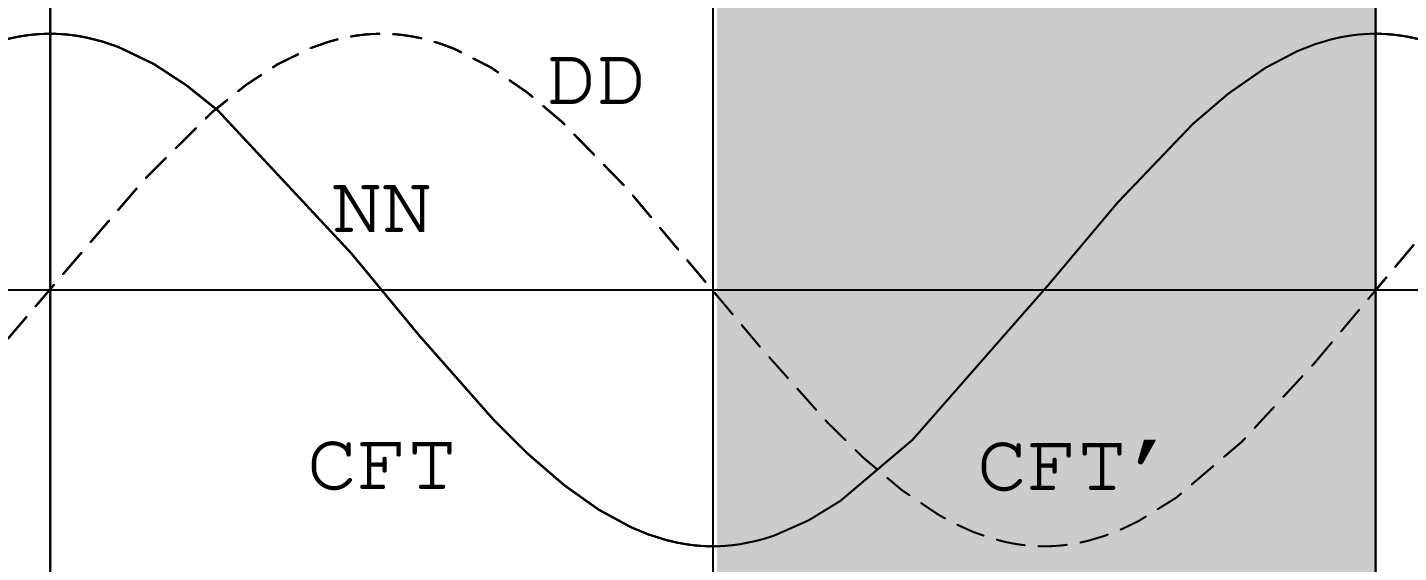}%
\qquad\qquad
  \includegraphics[height=2.5cm]{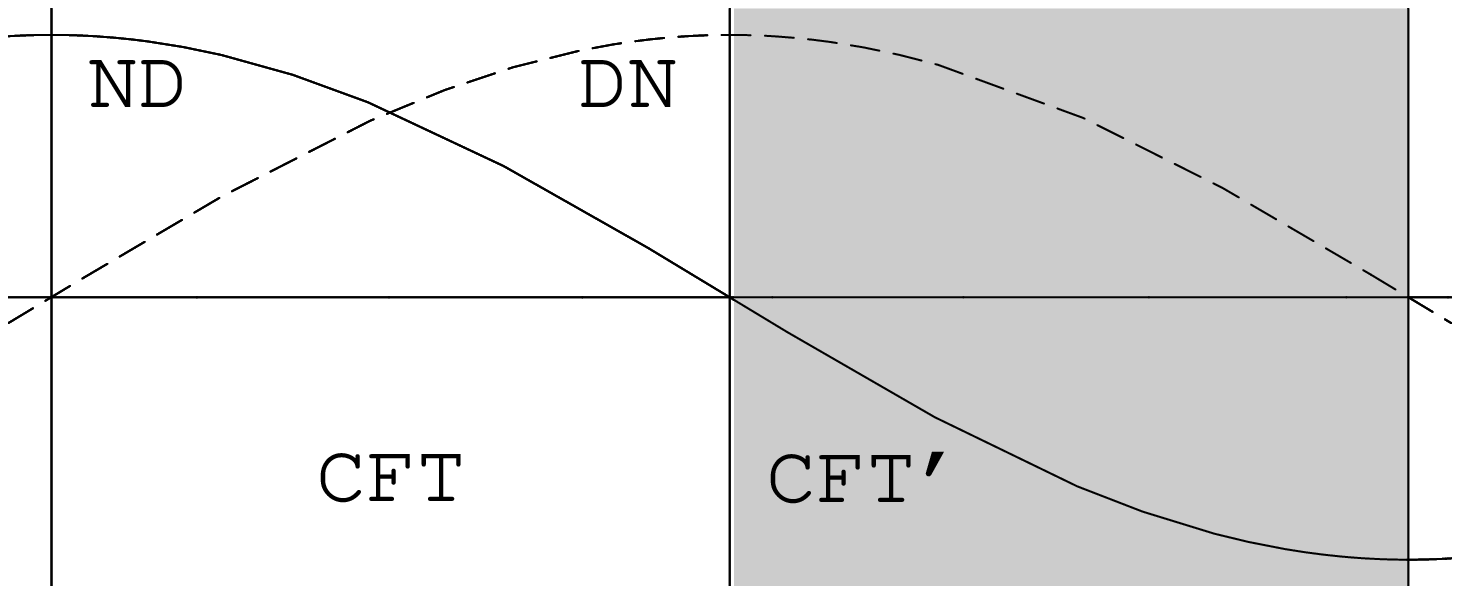}%
\end{center}
\caption{First excited modes along the circle in $AdS_3\times S_1$
for periodic (left) and antiperiodic (right) boundary conditions.
The unshaded (shaded) region represent the spaces where the CFT
(CFT$'$) are defined, both of them of the form $AdS_3 \times I$. }
\label{fig:reflecting} %
\end{figure}

The reason why this can be viewed as reflecting boundary
conditions (for the purpose of computing $\tmn$) is the following.
As depicted in Fig. \ref{fig:reflecting}, we can always expand the
modes on the $S_1$ in terms of sines and cosines and choosing
appropriately the phase, we have modes that are either Neumann (N)
or Dirichlet D at the two equatorial points of the $S_1$ (which
coincide with the boundary of $I$). Schematically, a periodic
field can be viewed as a sum of direct products of NN and DD
fields:
$$
\psi_{S_1}^{P} = {1\over \sqrt2}\lp(\phi_{I}^{NN} \otimes
{\phi'}_{I'}^{NN} + \phi_{I}^{DD} \otimes {\phi'}_{I'}^{DD} \rp)~.
$$
Similarly, for the antiperiodic fields one as
$$
\psi_{S_1}^{AP} = {1\over \sqrt2}\lp( \phi_{I}^{ND} \otimes
{\phi'}_{I'}^{DN} + \phi_{I}^{DN} \otimes {\phi'}_{I'}^{ND} \rp)~.
$$
These superpositions are nothing but the one noted in
\cite{massimoHiggs}, where it was argued that the KR boundary
conditions correspond to a mixture with equal weights of the modes
of $R\times S_3$ that are either symmetric or antisymmetric under
the reflection that maps one hemisphere of the $S_3$ into the
other. Indeed, in our notation the NN$\times$NN and DN$\times$ND
modes are symmetric while the DD$\times$DD and ND$\times$DN modes
are antisymmetric.

The main lesson of \cite{massGaugeInv} is that a superposition of
modes such as above can give the graviton a mass. The key element
of the computation is a crossed term in the graviton self-energy
$\sim \langle T_{\mu\nu}T_{\rho\sigma} \rangle$ which vanishes
unless both symmetric and antisymmetric modes are present.
However, here we shall only compute the one point function $\tmn$,
and this cannot contain crossed terms (at the quadratic level,
that is in the 1-loop approximation). In other words, the $\tmn$
for KR (anti)periodic boundary conditions and for the appropriate
combination of D and N conditions on $I$ should be the same. More
precisely, one should have
$$
P_{0\,(S_1)}^{P}=P_{0\,(I)}^{NN}+P_{0\,(I)}^{DD} %
\qquad {\rm and} \qquad %
P_{0\,(S_1)}^{AP}=P_{0\,(I)}^{ND}+P_{0\,(I)}^{DN}~.
$$
This equivalence can of course be broken when the CFT is
interacting, so in that case by `reflecting' boundary conditions
we will just mean the KR ones with the choice \eqref{reflecting}.

%
%

Finally, notice an important difference that we can already
foresee between the Casimir energies with KR and reflecting
boundary conditions, Eqs. \eqref{P0_KR} and \eqref{P0_refl}.
Since, as we will see, $P_0(\Delta\theta)$ has a simple zero at
$\Delta\theta=2\pi$ and for small DW tension
$\Delta\eta=\pi+O(\sigma)$, one automatically obtains that for
small $\sigma$
$$
P_0^{refl}\simeq 2 P_0^{KR}~,
$$
which does not seem surprising after all. This turns of to be of
some relevance for the interpretation of our results in terms of
massive gravity, see  Section \ref{sec:massgrav}.

\subsubsection{Casimir Energy on $AdS_3\times S_1$ at weak coupling}
\label{sec:ECas_weak}

Let us now evaluate the Casimir energy on  $AdS_3\times S_1$ at
one loop for arbitrary conformally coupled fields.
This can be easily achieved by exploiting the conformal invariance
of the field theory and the conformal properties of the
background.

Writing the AdS${}_3$ line element in horospheric coordinates
$ds_{AdS_3}^2=(dz^2+dx^2-dt^2)/z^2$, one readily sees that the
metric \eqref{s1ads3} is conformal to the  spacetime created by a
Cosmic String (CS)
\begin{equation}\label{cs}
z^2 d\theta^2+\, dz^2+dx^2-dt^2~, 
\end{equation} %
which has a deficit angle $2\pi-\Delta\theta$. The Casimir energy
density in this spacetime has been widely studied
\cite{helliwell,fsCS,dowkerCS,dowkerCS_Spin} and these results can
easily be transformed to the case of our interest. Since in the CS
space the circles defined by $\theta$ are contractible, in these
computations the fermions obey antiperiodic boundary conditions,
which is what we need.

For the CS spacetime, the anomaly term vanishes  and $\tmn$ is of
the form
\begin{equation}
    \tmn^{(CS)}={P_{0}^{(CS)} \over z^4}\, {\rm diag}
    \big(\frac{1}{3},-\frac{1}{3},-\frac{1}{3},1\big)_{\mu\nu}\,.
    \label{TmnCS}
\end{equation}
Using again \eqref{cftf}, one realizes that $P_{0}^{(CS)}=P_{0}$
so we can straightforwardly identify the Casimir energy.
For a collection of $N_0$ conformal scalars, $N_{1/2}$ Majorana
fermions and $N_1$ vectors, the result is \cite{fsCS}
\begin{equation}\label{spins}
P_0^{1-loop}={1-\nu^2\over 1920\pi^2}\lp[4 N_0 (\nu^2+1)+  N_{1/2}
(7 \nu^2+17)+8 N_{1} (\nu^2+11)\rp]
\end{equation}
where $\nu=2\pi/\Delta\theta$.

For $\mathcal{N}=4$ Super-Yang-Mills (SYM) with $N$ colours, it is
convenient to introduce the rescaled Casimir energy
\begin{equation}\label{p}
  p\equiv {32\pi^2 \over 3 N^2} \; P_0 ~,
\end{equation}
which measures the Casimir energy per colour degree of freedom.
The field content is $N_0=6N^2$, $N_{1/2}=4N^2$ and $N_1=N^2$, so
one finds\footnote{The result \eqref{ECas1loop} can also be
obtained by analytical continuation of the thermal state of the
CFT on the Hyperbolic plane \cite{emparan} with temperature
$1/\Delta\theta$ (which is dual to the hyperbolic Schwarzschild
AdS${}_5$ black Hole \cite{emparan}). This is for granted because
both configurations have the same Euclidean section.}
\begin{equation}\label{ECas1loop}%
p^{1-loop}(\Delta\theta)=
\lp(1-\lp[{2\pi\over\Delta\theta}\rp]^2\rp)\lp(1+
{1\over3}\lp[{2\pi\over\Delta\theta}\rp]^2\rp)~.
\end{equation}
As advanced, this vanishes for $\Delta\theta=2\pi$ (it does so for
each spin). For small $\Delta\theta$, it behaves as a usual
Casimir force  $ \sim\,1/\Delta\theta^4$.

\begin{figure}[tb]
\begin{center}
\includegraphics[height=4.5cm]{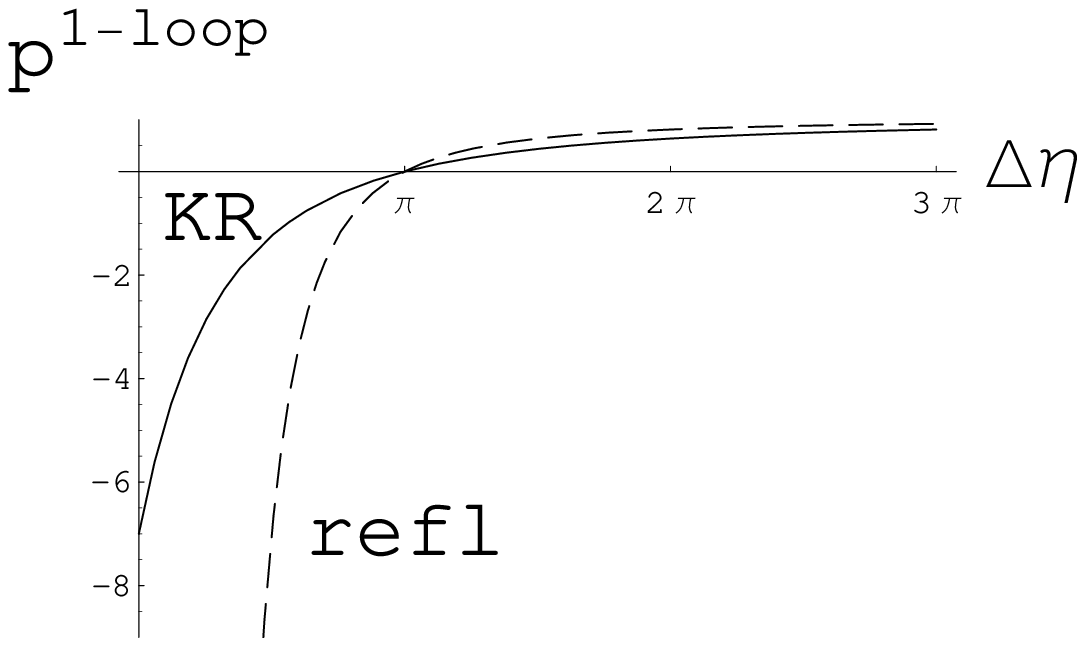}%
\qquad%
\includegraphics[height=4.5cm]{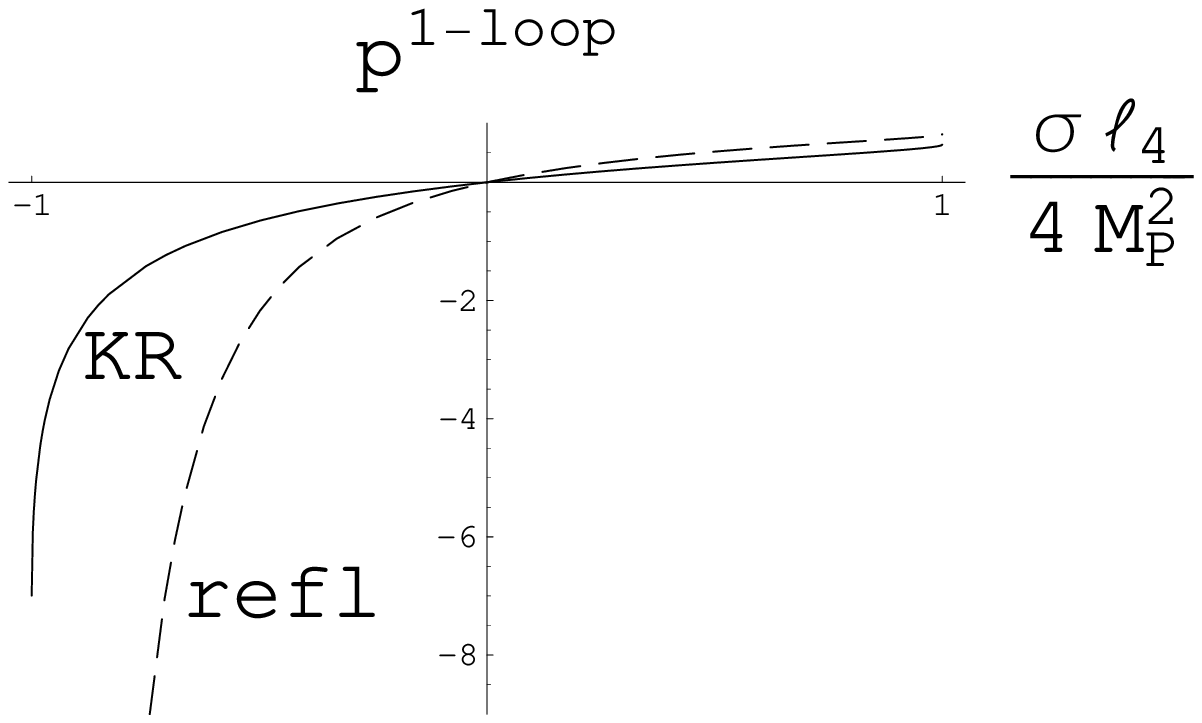}%
\end{center}
\caption{ Left panel: Casimir energy (density) for the CFT as a
function of the conformal interval $\Delta\eta$ for KR boundary
conditions (solid line) and for reflecting boundary conditions
(dashed line). These results are obtained from \eqref{ECas1loop}
taking $\Delta\theta=\pi+\Delta\eta$ or $\Delta\theta=2\Delta\eta$
for KR or reflecting boundary conditions respectively.
Right panel: The same, as a function of the DW tension $\sigma$
or, equivalently, its acceleration. Note that for small $\sigma$
the Casimir energy with reflecting boundary conditions is twice as
much that with KR conditions.
%
}
\label{fig:p1L} %
\end{figure}

Once we know $p^{1-loop}(\Delta\theta)$, we can obtain it as a
function of the conformal interval $\Delta\eta$ for KR and
reflecting boundary conditions by means of \eqref{P0_KR} and
\eqref{P0_refl} respectively. Also, since $\Delta\eta$ depends on
the DW tension $\sigma$ through \eqref{Deta}, it is also
straightforward to obtain $p^{1-loop}$ as a function of $\sigma$
or alternatively its acceleration. These results are plotted in
Fig. \ref{fig:p1L}.

%

Thus, we find that the vacuum polarization induced by subcritical
walls (may it be viewed just as a Casimir effect or as a cold bath
of radiation created by a mirror with acceleration $K_0<\ell_4$)
is, as expected,  nonzero. As we will explain in Section
\ref{sec:CS}, the phenomenon dual to this in the gravity side is
the fact that pure-tension codimension-2 branes generate a long
range Weyl curvature (aside from a deficit angle) in $AdS$ space.

It is also possible to compute $\p(\Delta\theta)$ at strong 't
Hooft coupling (still ignoring the backreaction)
using the `standard' methods of AdS/CFT, that is, without
dynamical gravity in the 4D side. We will do this in Section
\ref{sec:ECas_strong}, but let us advance that the difference with
respect to the weak coupling result \eqref{ECas1loop} is quite
small (see Fig \ref{fig:T_p_ratios}). Hence, we conclude that
strong coupling does not prevent the DW from radiating CFT quanta
(on the contrary, we will find that for positive DW tension, the
amount of radiation is actually \emph{larger} for
$\lambda\to\infty$). One may wonder whether this conclusion can be
changed once we introduce the backreaction. As we shall soon see,
it will not.

\subsection{Including the Backreaction}
\label{sec:Backreact}

Let us now consider the full problem, since the stress tensor that
we have just computed (see Eqs. \eqref{statedep.T}, \eqref{P} and
\eqref{p}) will feed back and modify the geometry that we assumed
\eqref{Rcases}. Our goal is to self-consistently solve the
semiclassical Einstein's equations
\begin{equation}\label{semiclassical}
  \mpl^2 \, G_{\mu\nu}= - \Lambda_4 \,g_{\mu\nu}+ T_{\mu\nu}^{DW} +
  \tmn^{CFT}~.
\end{equation}
In the thin wall approximation, we can separate the problem by
first working out the exterior of the DW and then gluing the two
sides by means of an appropriate matching condition.
Away from the DW, the semiclassical equations
\eqref{semiclassical} reduce to a `Friedman' equation, that can be
conveniently written as \cite{gripu}\footnote{The second piece in
the r.h.s. of \eqref{friedman} is the anomaly term. Here, we
assume that there is no contribution from the counterterm
$\sqrt{-g}R^2$ as it breaks conformal invariance.}
\begin{equation}
\label{friedman}%
\frac{R'^{\,2} - \kappa}{R^{\,2}}={1\over \ell_4^2}-
\frac{\ell_5^2}{4}
    \left(\frac{R'^{\,2} - \kappa}{R^{\,2}}\right)^2+%
    {\ell_5^2\over 4 }
  \,{\p(\Delta\eta)\over R^4}
\end{equation}
where $\p$ is defined in \eqref{p}
and we introduced
\begin{equation}\label{ell}
\ell^2_5 = {N^2 \over 8\pi^2\mpl^2}~.
\end{equation}
As is obvious from \eqref{friedman}, $\ell_5$ is the scale that
controls the backreaction and because of the large $N$ limit
assumed, it is much larger than the Planck scale. In fact,
$\ell_5^2$ is the gravitational version of the 't Hooft coupling,
and is the relevant scale because it is the combination that
enters in the CFT loops. This is also one reason why it plays the
role of the inverse cutoff of the theory \cite{ahpr,gia07,dr}.
Needless to say, in the gravity dual $\ell_5$ is the curvature
radius of $AdS_5$.

On the other hand, the matching condition also receives a
correction from the anomaly \cite{gripu},
\begin{equation}
\label{loop.corr}%
K_{0}\lp(1+\frac{1}{6}\, \ell_5^2 \lp( K_{0}^2-{3\kappa\over
R_{0}^2}\rp)\rp) =\frac{\sigma}{4\mpl^2}
\end{equation}
where as before $K_{0}=-(R'/R)_{0+}$ is the extrinsic curvature of
the DW.

Let us emphasize that Eqs. \eqref{friedman} and \eqref{loop.corr}
describe the whole CFT + gravity system coupled to the DW also
beyond the 1-loop or the planar approximations (in the maximally
symmetric configuration). Indeed, the form \eqref{statedep.T} of
the stress tensor is fixed by the symmetries and conformal
invariance, and the conformal anomaly does not receive higher loop
corrections for ${\cal N}=4$ SYM. The dependence on the 't Hooft
coupling $\lambda=g^2_{YM}N$ and the corrections from non-planar
diagrams enter only in the functional form of $\p(\Delta\eta)$. As
we have seen, for subcritical walls this is given by
\eqref{ECas1loop}, \eqref{general} at $\lambda=0$. In Section
\ref{sec:ECas_strong}, we shall obtain the non-perturbative form
of $\p$ at strong coupling ($\lambda\to\infty$) to leading order
in $1/N$ using `standard' AdS/CFT technology. As we will see, they
differ very little (see Eq. \eqref{p0NP} or Fig.
\ref{fig:T_p_ratios}), so to fix ideas we shall stick to the weak
coupling expression \eqref{ECas1loop}.

Hence, for every given value of $\lambda$ (and of $N$, in
general), $\p(\Delta\eta)$ is a fixed function that depends on
the geometry only through $\Delta\eta$. %
Then, one way to find the self-consistent solutions of
\eqref{friedman} and \eqref{loop.corr} in general is as follows.
We substitute momentarily $\p^{1-loop}(\Delta\eta)$ by a constant,
call it $p_*$. Then, it is straightforward to integrate
\eqref{friedman}, \eqref{loop.corr} and the resulting warp factor
$R_*(y)$ depends parameterically on $p_*$, as well as on the
dimensionless quantities
$$
\epsilon\equiv \lp({\ell_5\over \ell_4}\rp)^2
$$
(this is the relevant dimensionless gravitational coupling
constant) and
$$
\wt\sigma\equiv{\sigma\ell_4\over4\mpl^2}~.
$$
Then, we compute the conformal interval
$\Delta\eta_*(p_*;\epsilon,\wt\sigma)$ for $R_*(y)$,
\begin{equation}\label{Deta*}
  \Delta\eta_*(p_*;\epsilon,\wt\sigma)\equiv 2
  \lp[\int_{R_b}^\infty+\,{\rm sign}\sigma \int_{R_b}^{R_0}\rp]
{1\over R}{dR \over R'(R)}
\end{equation}
where $R_b$ is the warp factor at the bounce and $R'(R)$ is
obtained from \eqref{friedman}.
Then, the self-consistent solutions at 1-loop in the conformal
fields must satisfy
\begin{equation}\label{self}
\Delta\eta_*\lp[\p^{1-loop}(\Delta\eta);\,\epsilon,\,\wt\sigma\rp]=
\Delta\eta~.
\end{equation}
This defines the relation between $\Delta\eta$ and the tension
that fully incorporates the backreaction (it is built-in that it
reduces to \eqref{Deta} for $\epsilon=0$).%
\footnote{Needless to say, the extension of \eqref{self} for any
value of $\lambda$ is just given by the same relation but with the
corresponding form of the Casimir energy
$\p^{\lambda}(\Delta\eta)$ as the first argument in the r.h.s.}
Note that combining \eqref{self} and \eqref{general}, the same
condition can be cast as
\begin{equation}\label{self2}
  \Delta \theta^{1-loop}(\p) =
  \Delta\eta_*\lp[\p\,;\,\epsilon,\,\wt\sigma\rp]+\Delta\eta'~,
\end{equation}
where $\Delta \theta^{1-loop}(\p)$ denotes the inverse of
\eqref{ECas1loop}, and it is transparent that this determines the
Casimir energy as a function of the tension, the backreaction
parameter and the boundary conditions,
$\p=\p(\wt\sigma,\epsilon,\Delta\eta')$ (since the l.h.s. is a
function of $p$ only). In the form \eqref{self2}, the
self-consistency condition has a straightforward geometrical
interpretation in the 5D dual.

This procedure works for values of $\ell_5$ (in principle)
arbitrarily large, so one could keep track of the backreaction to
all orders in $\ell_5$.
Here, though, we shall not pursue this task because $\ell_5$ is
the inverse cutoff, so the only trustable solutions are those
where $\ell_5$ is small as compared to any other scale and we
shall content ourselves with the backreaction from $\tmn$ on the
geometry to the leading order, {\em i.e.}, $O(\ell_5^2)$.
Note that this will not prevent us from finding new solutions that
exist only because of the backreaction and which still lie within
the validity of the effective theory. These solutions necessarily
have no classical limit (typically when sending $\ell_5\to0$ the
curvature scales grow beyond $1/\ell_5$), and we shall call them
`Bootstrap' Domain Wall spacetimes. Examples of these solutions
will be discussed in Section
\eqref{sec:qs}. \\

Before addressing the DW spacetimes, let us briefly see what
happens in the absence of any DWs. An important question that we
can already answer is whether in the absence of DWs, the solution
of \eqref{friedman}, \eqref{loop.corr} is unique. As it turns out,
this is not entirely trivial and in fact strongly depends on the
form of the Casimir term $\p(\Delta\eta)$, that is, on the field
theory and on the boundary conditions. It is easy to check that
for ${\cal N}=4$ SYM and KR boundary conditions with
$\Delta\eta'=\pi$ (in the $\kappa=-1$ case), only pure $AdS$ (that
is, $\Delta\eta=\pi$, or equivalently $\p=0$) is a solution of
equation \eqref{self}. (Strictly speaking, there are solutions
other than $\p=0$ but only if $\ell_5\gtrsim\ell_4$, so they
cannot be trusted.\footnote{One may wonder if this conclusion
holds for any free conformal theory. With arbitrary $N_0$,
$N_{1/2}$ and $N_1$, the anomaly will contain in general a $\Box
R$ term (which would render \eqref{friedman} higher derivative)
but this can always be cancelled by appropriately choosing the
$\sqrt{-g}\,R^2$ counterterm. With this choice, it is also
possible to show that there is no choice of $N_0$, $N_{1/2}$ and
$N_1$ that generates self-consistent solutions other than $\p=0$
(with small $\epsilon$.)} In the gravity side these solutions will
not present, of course.) For reflecting boundary conditions one
has a similar situation.

\subsubsection{Massive gravity interpretation}
\label{sec:massgrav}

Going back to the DW case, let us see in full detail how the
presence of $\tmn$ affects the geometry to leading order.
Given that in the presence of a subcritical DW the Casimir energy
is already non-zero at zeroth order and that this enters in
\eqref{friedman} suppressed by $\ell_5^2$, we only need to find
the modification of the warp factor $R(y)$ to order $\ell_5^2$ and
this will represent the self-consistent solution of
\eqref{friedman} to leading order in $\ell_5^2$.
In the process, we shall give the massive gravity interpretation
of these solutions in terms of a particular pattern of screening
of the DW tension.

Already from the form of Eqs. \eqref{friedman} and
\eqref{loop.corr}, it is clear that generically some screening
takes place at two distinct levels. The first comes from the
boundary condition on the DW and so is local. The anomalous
contribution acts so as to effectively renormalize the DW tension
\cite{gripu}. We can define the \emph{effective} tension that is
generating the acceleration scale $K_0$ close to the wall
\eqref{loop.corr} as $\sigma+\delta\sigma_{local} \equiv 4 \mpl^2
K_0$. We shall use as a measure of the screening the ratio
$\delta\sigma/\sigma$.
Then, substituting the zeroth order values of $K_0$ and $R_0$ in
the $O(\ell_5^2)$ terms of \eqref{loop.corr} one obtains
\begin{equation}\label{delsigmaLoc}
  {\delta\sigma_{local}\over\sigma}=-\frac{\epsilon }{6}\lp( 3-2
  \,\wt\sigma^2\rp)~,
\end{equation}
which is negative for subcritical walls.
%
Hence, the acceleration produced by the DW is actually smaller
than the classical value \eqref{K0}. This looks like a genuine
screening phenomenon, similar to that found in \cite{dgpr} for the
DGP model \cite{dgp}. However, it is dubious that we should
ascribe it to the fact that the graviton is massive in this model
because it is independent of the boundary conditions chosen for
the CFT.\footnote{Besides, this effect is also present even for
$\Lambda_4\geq0$. However, it is intriguing that in these cases
the $3$ in \eqref{delsigmaLoc} is replaced by $0$ or $-3$, thus
obtaining \emph{anti}-screening instead.}
%
%
Rather, this kind of screening is associated to the fact that the
trace anomaly captures some of the short-distance properties of
the 5D dual. In terms of the gravity dual, this screening arises
because the DW is \emph{also} sourcing a deficit angle in the bulk
\cite{gripu}.

However, this is not the whole story because the Casimir energy
can also contribute to change the effective tension as measured
far away from the DW. It is not immediately obvious to us how to
rigorously define the notion of the DW tension measured at
infinity, so in the following we will resort to a convenient
coordinate system where the identification of the effective
tension at infinity seems natural.
%
This is given by a `Schwarzschild'-like coordinate, where the
metric is
\begin{equation}\label{Fmetric}
ds^2_4=\ell_4^2\lp( {dz^2\over F(z)}+F(z) \; ds_{AdS_3}^2\rp)
\end{equation}
with $F\equiv(R/\ell_4)^2$. In this gauge, the `classical'
background \eqref{Rcases} (with $\kappa=-1$) is
\begin{equation}\label{Fc}
F_c(z)=1+(|z|-z_0)^2
\end{equation}
where
\begin{equation}
z_0\equiv {\wt\sigma\over \sqrt{1-\wt\sigma^2}}~.
\end{equation}
The advantage of this gauge is that the different contributions to
the Newtonian potential $(F-1)$ are simply added up. Expanding
\eqref{Fc}, one identifies the usual quadratic potential due to
the cosmological constant, and a the linear potential $-2 z_0 |z|$
generated by the wall at $z=0$. In terms of this metric potential,
then, a screening of the DW tension translates into a modification
of the coefficient of the term linear in $z$.

To leading order in the backreaction the solution can be separated
as
$$
F(z)=F_c(z)+\delta F(z)
$$
where $\delta F$ is of order $\epsilon$. Introducing this
expansion in the equations of motion \eqref{friedman},
\eqref{loop.corr} and isolating the $O(\epsilon)$ terms leads to
the following equations for $\delta F$
\begin{eqnarray}
  {1\over2} {F'_{c}\over F_c} \, \delta F'-{\delta F \over \,F_c }
  &=&
{\epsilon\over 4}
\, \lp(  {\p\over F_c^2} - 1 \rp)\\
 \lp({\delta F'\over F_c'} -{1\over2}{\delta F \over F_{c}} \rp)\Big|_{z=0^+}
 &=& {\epsilon \over 2} 
 \, \lp(1 -{2\over3} \wt\sigma^2 \rp)
\end{eqnarray}
where a prime denotes differentiation with respect to $z$. The
solution is
\begin{eqnarray}
\delta F&=&{\epsilon\over 12}
\,\Big\{3-(z-z_0)(3z-(6+z_0^2)z_0) \cr %
&& \qquad\qquad%
-3 \p\;\lp[1-(z-z_0)(\arctan(z-z_0)+\arctan{z_0})\rp]
 \Big\}
\end{eqnarray}
Expanding this for $z\to\infty$, one identifies that the effective
tension perceived at infinity differs from $\sigma$ by
\begin{equation} \label{delsigmaTot}
  {\delta\sigma_\infty \over \sigma}=
  {(\delta z_0)_\infty \over \wt\sigma {dz_0\over d\wt\sigma}}=%
  \epsilon\,\lp\{{1\over8}-{3-2\,\wt\sigma^2\over6} + %
  {(1-\,\wt\sigma^2)^{3/2}\over 16 \, \wt\sigma} \lp(\pi +2 \arctan{\wt\sigma\over\sqrt{1-\wt\sigma^2}} \rp)
  \,
  \p \rp\}.
\end{equation}
The first contribution in the curly brackets comes from the
anomaly term away from the DW which, even for $\p=0$, renormalizes
the $AdS$ curvature $\ell_4^{-2}\to (1-\epsilon/4) \ell_4^{-2}$.
The second contribution is precisely the local screening term
\eqref{delsigmaLoc} that takes place at the DW location itself as
we discussed.
The remaining $\p$-dependent term is the truly non-local effect,
due to the Casimir stress.

\begin{figure}[t]
\begin{center}
  \includegraphics[height=5cm]{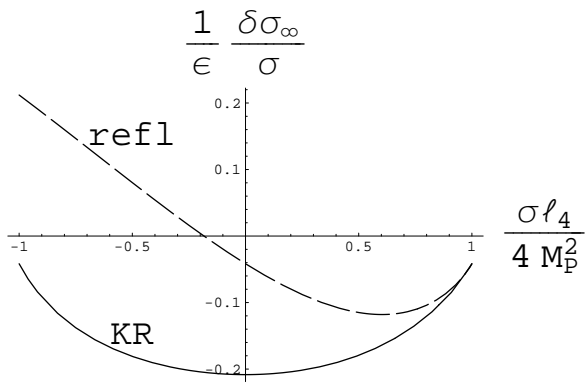}%
\end{center}
\caption{The amount of screening of the DW tension measured at
infinity, Eq. \eqref{delsigmaTot}, enhanced by a factor
$1/\epsilon$. Clearly, KR boundary conditions give more screening
($\delta\sigma_\infty/\sigma$ is more negative) than the
reflecting ones. }
\label{fig:screening} %
\end{figure}


We plot the total effect at infinity $\delta\sigma_\infty /\sigma$
both for KR and reflecting boundary conditions in Fig.
\ref{fig:screening}. It is encouraging to see that for KR
conditions one always has screening ($\delta\sigma_\infty
/\sigma<0$). For reflecting conditions, one can have either
anti-screening or screening, depending on the value of $\sigma$.
Furthermore, it is safe to say that one has always \emph{more}
screening with the KR choice (that is, $\delta\sigma_\infty
/\sigma$ is more negative in this case). It is of course quite
tempting to ascribe this phenomenon to the fact that the graviton
is massive with the KR conditions.

The reason why $\delta\sigma_\infty /\sigma$ is more negative for
KR conditions can be traced back to the form of the Casimir energy
$p(\sigma)$. In particular, it relies on 1) $p$ is larger (in
magnitude) for reflecting boundary conditions and 2) the sign of
$p$ is the same as that of $\sigma$. Thus, in any field theory
where these two properties hold, KR boundary conditions will give
more screening.

Let us finish by mentioning that even though for KR conditions the
graviton is massive
\cite{kr,massGaugeInv,massimoHiggs,massimoReview,duffLiuSati,giannakisLiuRen}
with a mass $m_g^2\sim \epsilon / \ell_4^2$, it seems that (for
the DWs) we do not find any trace of a `Compton wavelength' scale
in the self-consistent solution.
Indeed, the radiation `halo' is peaked around the bounce with a
typical (proper) width of order $\ell_4$, as pictured in  Fig.
\ref{fig:picture}. Hence, the largest modification of the geometry
occurs in this region%
, and
it is natural to expect that the non-local screening basically
occurs there. However, the bounce is a proper distance
\begin{equation}\label{y*}
y_*\simeq \ell_4 \,{\rm arctanh}\lp({\sigma\ell_4\over
4\mpl^2}\rp) + {\cal O}(\epsilon)
\end{equation}
from the wall. This scale has little to do neither with the
Compton wavelength $\ell_4/\sqrt\epsilon$ nor the scale at which
the deviations from massless $AdS$ gravity $\sim \,
\ell_4/\epsilon$ were found for the shock wave solutions
\cite{ks}. The distance between the wall and the (center of the)
radiation cloud, $y_*$, is larger than the width of the radiation
cloud itself only for walls already quite close to critical, and
it becomes larger than $m_g^{-1}$ (as in the shock wave case) only
for tensions exponentially close to critical. For moderate
tensions, though,  all forms of the screening take place within
one curvature radius, $\sim\ell_4$.

Hence, at least for the DWs, the graviton mass does not manifest
itself by displaying a different behaviour in the metric at a
certain length scale related to the mass. This might be due to the
large amount of symmetry that we have assumed, and hence might be
only a particular feature of the maximally symmetric DWs. Instead,
the presence of a larger screening for KR boundary conditions as
summarized in Fig. \ref{fig:screening} seems a more clear
indication that gravity is massive in this case.


\subsubsection{Bootstrap Domain Wall Spacetimes} \label{sec:qs}

Let us now discuss a new kind of asymptotically $AdS$ solutions
that that appear thanks to the backreaction from the Casimir
energy itself. These solutions have no classical counterpart, and
we will call them `Bootstrap DW spacetimes'. As we will see, some
of these solutions (those with a positive tension DW) display a
bounce that is supported by the Casimir energy.
%
%

In these solutions, the link between whether the tension is sub-
or super-critical and the sign of the curvature of the DW
worldsheet ($\kappa$) is lost, simply because what determines
$\kappa$ is now the balance between the DW tension against a
combination of $\Lambda_4$ and the Casimir energy density next to
the DW. Hence, from now on, we shall call the solutions with
$\kappa=0$ `planar' and those with $\kappa=1$ `inflating'.


As argued in \cite{gripu}, the reason why such solutions do not
exist with $\Lambda_4\geq0$ is that in order that the curvature
scale at the bounce is below the cutoff, $\p$ must be large and
negative. As we have seen in Section \ref{sec:ECas_weak}, this can
be accomplished if the conformal interval $\Delta\eta$ is small
enough. But this is impossible in an asymptotically flat or de
Sitter space (because $R(y)$ does not grow fast enough at
infinity), so the only solutions of this form have the direction
transverse to the DW compact (with a finite \emph{proper} length),
and would represent a DW in a `cage'. For $\Lambda_4<0$, the
situation is very different because $AdS$ acts as the cage. In
other words, a finite conformal interval does not imply that the
space is compact in this case.

Going back to $\Lambda_4<0$,  from \eqref{friedman}, the bounce
occurs at
\begin{equation}\label{Rb}
R_b^2={\ell_4^2\over2}\lp(-\kappa+\sqrt{(1+\epsilon)\kappa^2-\epsilon
\p}\rp)~.
\end{equation} %
For inflating walls, this is only larger than the cutoff
$\ell_5^2=\epsilon\ell_4^2$ if $|\p| \gtrsim 1/\epsilon$.
In these cases the solutions can be trusted.
For planar walls, \eqref{Rb} is larger than $\ell_5$ even if $\p$
is of order one, but this is only a manifestation that this case
represents the limit $\p\to-\infty$, $R_b\to\infty$ of the
$\kappa=1$ case. In fact, for $\kappa=0$ the value of $R(y)$ at a
given point has no meaning because it can be scaled away by a
change of coordinates. In this case, all curvature invariants are
of order $1/\ell_4$ irrespective of $\p$.\\

Let us consider the planar case ($\kappa=0$) in more detail (the
following discussion parallels in many respects that of \cite{gpt}
in a five dimensional context; similar arguments apply to
inflating walls, but we shall leave this case for the future).
As we shall now see, when the backreaction from the Casimir effect
is included, there is always a range of DW tensions for which
there are self-consistent solutions where the  DW is planar but
the tension is subcritical.

Since the radiation term in \eqref{friedman} is at most comparable
to $\Lambda_4$ (close to the bounce), to leading order in
$\epsilon$ we can safely neglect the anomaly term. Hence, the
Friedman equation reduces to
\begin{equation}
\label{friedman0}%
\frac{R'^{\,2}}{R^{\,2}}\simeq{1\over \ell_4^2}
    +{\ell_5^2\over 4 }
  \,{\p\over R^4} ~.
\end{equation}
Note that in the planar DW case, we can always rescale the
coordinates so that we can fix $R$ at will at one point. This
should translate in \eqref{friedman0} as a scaling symmetry $R\to
\gamma R$. The radiation term does have this symmetry because for
$\kappa=0$ there is only one scale ($\Delta\eta$), so by conformal
invariance the Casimir must take the form
$\p\sim1/(\Delta\eta)^4$, hence it scales like $\p\to \gamma^4
\p$.

The solution of \eqref{friedman0} is
\begin{equation}
  R(y)=R_b\,
  \cosh^{1/2}[(2|y|-y_0)/\ell_4]~,
\end{equation}
where now $ R_b= (-\epsilon \p/4)^{1/4}\,\ell_4$. The integration
constant $y_0$ is fixed by the junction condition at the DW, which
for planar walls reads
\begin{equation}
\label{junct.crit}%
K_{0}\lp(1+\frac{1}{6}\, \ell_5^2 K_{0}^2\rp)
={\wt\sigma\over\ell_4}
\end{equation}
Ignoring the anomalous correction, this gives the same form for
$y_0$ as in \eqref{y*}.
With this, the redshift factor between the DW and the bounce is
found to be
$$
{R_0\over R_b}\simeq {1\over (1-\wt\sigma^2)^{1/4}}~.
$$
Note that in order to obtain these planar solutions, what we are
doing is to tune the DW tension against $\Lambda_4$ plus the
Casimir term.

Yet, this does not make the solution self-consistent. For this, we
still have to solve Eq. \eqref{self}. The function $\Delta\eta_*$
can be easily obtained explicitly in this case,
\begin{equation}
\Delta\eta_*(p_*)={h(\wt\sigma)\over (-\epsilon \, p_*/ 4 )^{1/4}}
\end{equation}
where
\begin{equation}
h(\wt\sigma)=(1+{\rm sign}\,\wt\sigma) {2\sqrt\pi
\,\Gamma({5\over4})
\over \Gamma({3\over4})}%
-{1\over2}\,{\rm sign}\,\wt\sigma\;B_{(1-\wt\sigma^2)}\Big({1\over4},{1\over2}\Big)   %
\end{equation}
with $B_z(a,b)$ the incomplete Beta function. Note that
$h(\wt\sigma)$ vanishes for $\wt\sigma\to-1$, as it should since
in this limit the DW chops off all of the space.
%
%
%
%
%

Next, we need the form of $\p(\Delta\eta)$ at, say,  one loop for
the planar walls. With the same kind of Karch-Randall boundary
conditions, this boils down to computing the Casimir energy on
$S_1\times R^3$. Because there is no other scale in the problem,
$\p(\Delta\theta)$ can only be of the form $\Delta\theta^{-4}$,
where $\Delta\theta=\Delta\eta+\Delta\eta'$ is the length of the
circle. Since the result must match the $S_1 \times AdS_3$ case
for $\Delta\theta\to0$, it can only be
\begin{equation}\label{p0Planar}
\p^{1-loop}_{KR}=-{1\over3}\lp(2\pi\over\Delta\eta+\Delta\eta'\rp)^4
\end{equation}
where as before, $\Delta\eta'$ is the conformal interval in the
`adjacent' CFT$'$ and so is a constant that parameterizes the
boundary conditions. Note that it is not so clear that
$\Delta\eta'=\pi$ is the natural value now, because that does not
give global $AdS_4$ as the space where the CFT$'$ is defined
anyway. In the following, we shall assume a generic value for
$\Delta\eta'$.

The strong coupling result differs from \eqref{p0Planar} by a
factor $3/4$ (see Section \ref{sec:ECas_strong}). Of course, this
is the same factor that appears in the usual AdS/CFT computations
of the CFT at finite temperature
\cite{gubserKlebanovPeet,emparan}.
Hence, Eq. \eqref{self} reads
$$
\Delta\eta=h(\wt\sigma)\,\lp({4a(\lambda)\over\epsilon}\rp)^{1/4}{\Delta\eta+\Delta\eta'\over2\pi}
$$
where at weak and strong coupling we have $a(0)=3$ and
$a(\infty)=4$ respectively. Hence, we can express the conformal
interval directly in terms of the DW tension, the `t Hooft
coupling, the backreaction parameter $\epsilon$ and the boundary
conditions ($\Delta\eta'$) as
\begin{equation}\label{QuantumPlanar}
\Delta\eta=\Delta\eta'\; \lp[ {2\pi\over h(\wt\sigma)}\,
\lp(\epsilon\over4a(\lambda)\rp)^{1/4}-1 \rp]^{-1}~,
\end{equation} %
and the Casimir energy as
\begin{equation}
\p^{1-loop}_{KR}=-{1\over a(\lambda)}
\lp({2\pi\over\Delta\eta'}\rp)^4 \lp[1-{h(\wt\sigma)\over
2\pi}\lp({4 a(\lambda)\over \epsilon}\rp)^{1/4}\rp]^{4}
\end{equation}

\begin{figure}[t]
\begin{center}
  \includegraphics[height=5cm]{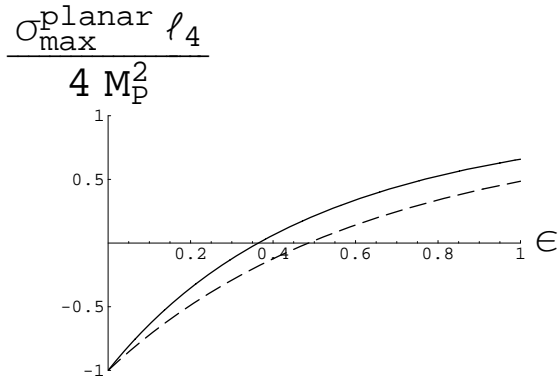}%
\end{center}
\caption{The planar Bootstrap DW solutions in the DW tension {\em
vs} $\epsilon $ diagram fall in the area below the curves (KR-like
boundary conditions are assumed). The solid and dashed lines
correspond to weak ($\lambda\to0$) and strong ($\lambda\to\infty$)
coupling results respectively. Even for small $\epsilon$, there
are solutions for a finite range of tensions close to the critical
value $-\sigma_c$ (see \eqref{sigmac}). The maximal tension
$\sigma^{planar}_{max}$ is smaller at strong coupling. Hence, one
can say that strong coupling effects in the CFT remove some of
these quantum solutions.}
\label{fig:planarQDW} %
\end{figure}

Thus, from \eqref{QuantumPlanar}  we see that self consistent
solutions exist whenever $\Delta\eta'\neq0$, and $h(\wt\sigma)$ is
small enough, that is, when the tension is close enough to
$-\sigma_c$, the critical value given in \eqref{sigmac}.
For every given $\epsilon\ll1$, the planar DW solution with
maximal tension $\sigma_{max}^{\,planar}$ is the one for which
$$
h(\wt\sigma_{max}^{\,planar})={2\pi}
\lp(\epsilon\over4a(\lambda)\rp)^{1/4}~.
$$
We plot how $\sigma_{max}^{\,planar}$ depends on $\epsilon$ in Fig
\ref{fig:planarQDW}. Note that, as mentioned above, these
solutions cease to exist in the $\epsilon\to0$ limit (only the
$\sigma=-\sigma_c$ solution would survive in this limit).
For small but finite $\epsilon$, though, there is a finite range
of the DW tension for which the solutions are trustable.
Naturally, this range starts up at $-\sigma_c$ because this is
when $\Delta\eta$ vanishes and hence $\p$ is maximal.
As Fig. \ref{fig:planarQDW} shows, for moderately small
$\epsilon\gtrsim1/2$ one can have planar solutions with positive
DW tension. Needless to say, at this point it is hard to tell
wether these solutions are not present in the underlying theory or
whether they are but with relative large corrections.

It is also worth pointing out that in principle there are
solutions with no DWs ($\sigma=0$) but still with some nontrivial
Casimir effect, which is entirely supported by the slightly
unconventional choice of boundary conditions, of course. These
arise for $\epsilon$ larger than
$$
%
\epsilon_c={4a(\lambda)\over\pi^2} \,\lp({\Gamma({5/4})\over
\Gamma({3/4})}\rp)^4%
$$
which is approximately $0.36$ ($0.48$) for weak (strong) coupling.
Hence, above this critical value new solutions of the gravity +
CFT system with this kind of asymptotic behaviour open up.

Finally, let us briefly mention what happens for reflecting
boundary conditions. In this case,
\begin{equation}\label{p0PlanarRefl}
\p^{1-loop}_{refl}=-{1\over3}\lp(\pi\over\Delta\eta\rp)^4
\end{equation}
and Eq. \eqref{self} directly gives a $\Delta\eta$-independent
equation that links the DW tension with  $\epsilon$ as
$$
h(\wt\sigma_{refl}^{\,planar})={\pi}
\lp(\epsilon\over4a(\lambda)\rp)^{1/4}~.
$$
Inverting this, one finds that the critical value of the tension
that gives a planar solution is very close to $-\sigma_c$ for any
$\epsilon<1$. Note, that both $\Delta\eta$ and $p$ are arbitrary
for this solutions. This degeneracy is expected to disappear in
the next order in $\epsilon$.

\section{5D Gravity dual}
\label{sec:5D}

In this Section, we discuss the 5D duals of the quantum corrected
DW solutions presented in Section \ref{sec:cft}. As before, we
shall first consider the gravity dual ignoring the backreaction on
the 4D metric, in Section \ref{sec:noBrane}. This is the
`standard' version of the AdS/CFT correspondence, and boils down
to finding the regular vacuum $AdS_5$ solutions (with no branes)
with a boundary metric conformal to the Domain Wall background. We
then holographically include dynamical 4D gravity in Section
\ref{sec:DWs in KR} by constructing the solutions where a DW
localized on a Karch-Randall brane.

\subsection{Ignoring the Backreaction (the brane)}
\label{sec:noBrane}

In Section \ref{sec:cft} we considered the maximally symmetric
configurations, where the DW worldvolume geometry is $M_\kappa$,
{\em i.e.},  a 2+1 dimensional de Sitter ($\kappa=1$), Minkowsi
($\kappa=0$) or Anti de Sitter space ($\kappa=-1$). Hence, we only
need to find asymptotically $AdS_5$ solutions with the same
symmetries. A generalization of the Birkhoff theorem guarantees
that the most general $\Lambda-$vacuum solution with these
symmetries can can be written locally as
\begin{eqnarray} \label{bulkmetric}%
ds_5^2&=&\ell_5^2 \,f(R) \,d\theta^2 + {dR^2\over f(R)}
+R^{2}ds_{\kappa}^2\,,\\%
f(R)&=&\kappa+{R^2\over\ell_5^2}+{\mu\over R^2} \nonumber
\end{eqnarray}
where $\ell_5^2=-6M^3/\Lambda_5$ is the AdS$_5$ curvature radius,
$M=(8\pi G_5)^{-1/3}$ is the bulk Planck mass and $\mu$ is an
integration constant proportional to the Weyl curvature of the
solution.
As before, $ds_\kappa^2$ is the line element on $M_\kappa$ with
unit radius.
%
The $\kappa=1$ case includes the (higher dimensional
generalization of the) BTZ black hole and the Schwarzschild-AdS
bubble of nothing solutions.
For $\kappa=0$ and $\mu\neq0$ \eqref{bulkmetric} is the
AdS-Soliton. 
Here we shall be most interested in the
$\kappa=-1$ case. For $\mu=0$, it gives global $AdS_5$ and for
$\mu\neq0$ it is what we will call a `hyperbolic AdS Solion'.

We shall be most interested in the `non-extremal' cases, for which
$f(R)$ has simple zeroes. Then, the larger root of $f(R)$ is
\begin{equation}\label{Rplus}
  R_+={\ell\over\sqrt2}\lp(-\kappa+\sqrt{\kappa^2-4\mu/\ell_5^2}\rp)^{1/2}
\end{equation}
Hence, $R$ ranges from $R_+$ to $\infty$ and $R_+$ represents the
center of the axial symmetry (which is present in the absence of
the brane and the DW).
In order not to have a conical singularity at $R=R_+$, one has to
periodically identify  $\theta$ with period
\begin{equation}\label{deltaTheta}
  \Delta \theta =
  {4\pi \over \ell_5 f'(R_+)}%
  =2\pi\,\lp({-\kappa+\sqrt{\kappa^2-4\mu/\ell_5^2}\over2\,(\kappa^2-4\mu/\ell_5^2)}\rp)^{1/2}~.
\end{equation}
Hence, in all these cases, the coordinate $\theta$ is periodic,
and the boundary is conformal to $S_1\times M_\kappa$. According
to the AdS/CFT correspondence, these solutions can be matched to
the states of ${\cal N}=4$ SYM on the boundary at strong coupling.
Let us now use this correspondence to infer the stress tensor of
the CFT at strong coupling for the case of our interest.

\begin{figure}[t]
\begin{center}
  \includegraphics[height=8cm]{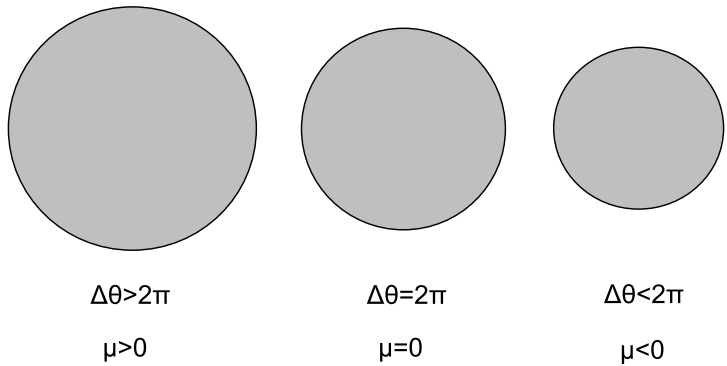}%
\end{center}\vspace{-2cm}
\caption{Hyperbolic $AdS$ solitons with different `radii'
$\Delta\theta$ can be represented as disks of different sizes.
Every point represents an $AdS_3$ with a curvature radius given by
\eqref{Rplus} at the origin while at infinity it grows as fast as
the circle spanned by $\theta$.
The boundary is of the form $S_1\times AdS_3$, and the length of
the $S_1$ in units of the $AdS_3$ radius is $\Delta\theta$,
\eqref{deltaTheta}. Global $AdS_5$ is the particular case
$\mu=0$.}
\label{fig:hyperSolitons} %
\end{figure}

\subsubsection{Casimir Energy on $AdS_3\times S_1$ at strong
coupling} \label{sec:ECas_strong}

For the $AdS_3$ slicing, the 5D metric corresponds to the
Hyperbolic $AdS$ Solitons, {\em i.e.}, Eq. \eqref{bulkmetric} with
$\kappa=-1$. In order that these solutions are regular, the
integration constant must obey $\mu < \ell^2_5/4$.

The structure of these spacetimes is very similar to global AdS.
To visualize it, it is convenient to suppress the $AdS_3$ factors.
The geometry spanned by the $R$ and $\theta$ `polar' coordinates
is a hyperboloid, with the topology of a disk. The boundary of the
disk has infinite volume, but we can always make a conformal
transformation to bring it to a finite size. Then, the difference
between these spaces \eqref{bulkmetric} with different values of
$\mu$ is that the length of this disk is different\footnote{The
Hyperbolic $AdS$ Solitons can actually be viewed as purely
gravitational Cosmic Strings (codimension 2 objects). We comment
on this in Section \ref{sec:CS}.}, as shown in Fig.
\ref{fig:hyperSolitons}.

\begin{figure}[t]
\begin{center}
  \includegraphics[height=5cm]{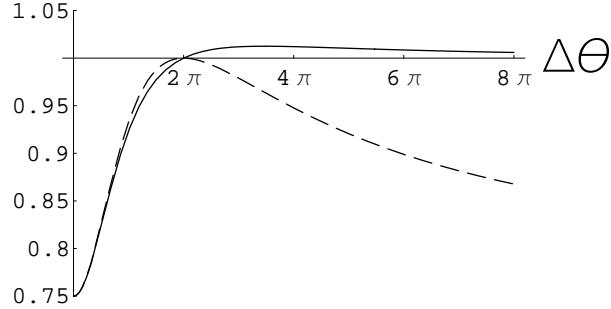}%
\end{center}\vspace{-.5cm}
\caption{Comparison between the weak and strong coupling forms of
the Casimir Energy on $S_1\times AdS_3$ with antiperiodic boundary
conditions for the fermions. The solid and dashed lines are
$\p^{NP}/\p^{1-loop}$ and $\langle T_0^0 \rangle^{NP}/\langle
T_0^0 \rangle^{1-loop}$ respectively. For a single DW and KR
boundary conditions $\Delta\theta$ ranges from $\pi$ to $3\pi$,
while for reflecting boundary conditions $0<\Delta\theta<4\pi$.}
\label{fig:T_p_ratios} %
\end{figure}

The important point is that the boundary of these geometries is of
the form $AdS_3\times S_1$ with different $S_1$ lengths.
In Section \ref{sec:ECas_weak}, we computed the 1-loop Casimir
energy of the CFT on this geometry. Now, using the AdS/CFT
correspondence, it is straightforward to obtain the
non-perturbative form of the Casimir energy for large 't Hooft
coupling to leading order in the $1/N$ expansion. According to the
correspondence, this is given by the energy of the 5D gravity
solution \eqref{bulkmetric}.
Since the metric \eqref{bulkmetric} is asymptotically $AdS_5$, one
can use the prescription introduced in \cite{balaKraus}.
Upon rescaling the boundary metric so that it coincides with
\eqref{s1ads3}, and expressing the 5D quantities directly in terms
of the CFT ($\ell_5^2 M^3 = N^2/4\pi$), one obtains
\cite{balaRoss}
\begin{equation}
    \tud^{NP}={3N^2\over 32\pi^2 }\, {1\over
R_*^4}\,\lp(1-{4\mu\over\ell_5^2} \rp)\;  {\rm diag}
\big(-\frac{1}{3},-\frac{1}{3},-\frac{1}{3},1\big)_\mu^{~\nu}\,.
    \label{Tstrong}
\end{equation}
The first piece in the parenthesis is readily identified as the
anomaly term. Thus, the second term is the non-perturbative form
of the Casimir energy, that is,  $\p^{NP}=4\mu/\ell_5^2 $.
Expressing $\mu$ in terms of $\Delta\theta$ by means of
\eqref{deltaTheta}, one obtains
%
\begin{equation}\label{p0NP}
\p^{NP}\lp(\Delta\theta\rp)=1-{2\pi^2\over\Delta\theta^2}-2\pi^4\,{1+\sqrt{1+2\Delta\theta^2/\pi^2}\over
\Delta\theta^4}~.
\end{equation}

This expression can be directly compared to the weak coupling
result \eqref{ECas1loop}. As shown in Fig. \ref{fig:T_p_ratios},
the difference between the weak and strong coupling Casimir
energies is quite modest. The maximum discrepancy is for
$\Delta\theta\to0$, for which one has the famous $3/4$ suppression
at strong coupling. It is noteworthy that for $\Delta\theta>2\pi$
the Casimir Energy at strong coupling is actually \emph{larger}
than at weak coupling, with a maximum enhancement by a factor
$81/80$.
Instead, the full stress tensor (including the anomaly term) is
always smaller for any $\Delta\theta$.
Notice as well that $\p^{NP}$ again vanishes for
$\Delta\theta=2\pi$ (pure $AdS_4$), as it should because $\mu=0$
is an exact string state.

Since, argued above, the Casimir energy can also be viewed as the
amount of particles produced by the accelerating DW, we conclude
that for this system strong coupling effects do not dramatically
suppress the particle production. As we shall see in Section
\ref{sec:DWs in KR} (together with the results of Section
\ref{sec:Backreact}), the inclusion of the backreaction does not
change this conclusion.

\subsection{Localized Domain Walls  in the Karch-Randall model}
\label{sec:DWs in KR}

Let us finally turn to the actual dual of the setup described in
Section \ref{sec:cft},  the Karch-Randall model \cite{kr}. The 4D
quantum corrected Domain Wall solutions simply correspond to the
solutions representing a Domain Wall localized on the brane (from
now on, by `the brane' we will mean the 3+1 brane, not the DW). In
the Karch-Randall model, the brane tension $\tau$ is set below the
critical or `Randall-Sundrum' value $\tau_{RS}=6 M^3 /\ell_5$ so
that its geometry is (asymptotically) $AdS_4$.

In the next derivation, we will follow Refs.
\cite{gp1,gripu,dgpr}. As before, we will concentrate on the
solutions with the symmetries of a maximally symmetric DW, that is
with a 3D maximally symmetric slicing. The (double-Wick rotated
version of the) Birkhoff theorem guarantees that the most generic
solution with this symmetry can be written locally as
\eqref{bulkmetric}.
In terms of these `bulk adapted' coordinates, the full spacetime
can be constructed as usual by finding the embedding of the brane
in the bulk, cutting the bulk along the brane and gluing two
copies by the brane location (we are assuming $Z_2$ symmetry
across the brane).

With the assumed symmetry, the location of the brane can be
parameterized by two functions $(R(y),\theta(y))$, and solve for
them by imposing that the Israel junction conditions are
satisfied. A level of arbitrariness is still present, due to the
re-parametrization (gauge) invariance of the embedding. To fix the
gauge, it is convenient to choose
\begin{equation}\label{gauge}
\ell_5^2 f(R)\theta'^2+{R'^2\over f(R)}=1\,.
\end{equation}
With this condition, the induced metric on the brane precisely
takes the form \eqref{metric}, and $y$ is the proper distance on
the brane orthogonal to the DW. Also, once $R(y)$ is known then
one can solve for $\theta(y)$ by integrating Eq. \eqref{gauge} and
the full solution will be determined.

As is well known, the Israel junction conditions for the brane
lead to a `Friedman' equation
\begin{equation}
\label{friedmanAdS}%
\frac{R'^2-\kappa}{R^2}=  \lp(1-{\epsilon\over4}\rp){1\over
\ell_4^2}
+ {\mu  \over R^4}%
 \,, 
\end{equation}
where now
$$
{1\over\ell_4^2}=-{\delta\tau\over3\mpl^2}
$$
(as before, $\epsilon\equiv \ell_5^2/\ell_4^2$) and we used
$M^3\ell_5=\mpl^2$.

Obviously, \eqref{friedmanAdS} is almost identical to the four
dimensional counterpart of the Friedman equation, \eqref{friedman}
once we identify the Casimir term as
\begin{equation}\label{p0_mu}
\mu={\ell_5^2\over4}\,\p~.
\end{equation}
To be precise, the brane-world version of the Friedman equation is
equivalent only to the leading order in the backreaction parameter
$\epsilon$, since to leading order the anomaly term in
\eqref{friedman} is just the constant $-\epsilon/(4\ell_4^2)$. It
immediately follows that the correspondence between the
brane-world setup (may it be the Randall-Sundrum or the
Karch-Randall model) and the 4D gravity + strongly coupled CFT
system holds to the leading order in the backreaction.

As in Section \ref{sec:cft}, the Friedmann equation
\eqref{friedmanAdS} holds away from the DW location, and in order
to take this into account one needs an appropriate matching
condition. As worked out in \cite{gripu}, this reads
\begin{equation}\label{junctionAdS} %
{1\over \ell_5 }\,\arctan\lp( {K_{0} \,\ell_5 \over
1-\epsilon/2}\rp) = {\sigma\,\over 4 \mpl^2}~,
\end{equation} %
where  $K_{0}\equiv -(R'/R)|_{0+}$, as before. This junction
condition is equivalent to \eqref{loop.corr} up to order
$\epsilon$, as expected because the leading correction in
\eqref{loop.corr} arises from the trace anomaly \cite{gripu}.

%
%

%
%
%
%

\begin{figure}[tb]
\begin{center}
  \includegraphics[height=9cm]{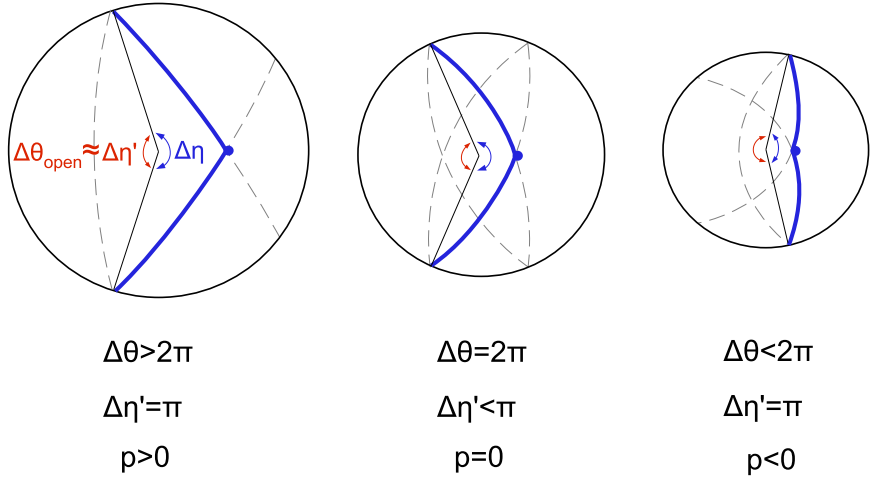}%
\end{center}\vspace{-3cm}
\caption{Schematic view of the localized Domain Walls in the one
brane case (corresponding to KR boundary conditions). The blue
solid line represents the 3+1 brane and the DW is the blue dot.
The bulk corresponds to two copies of the left side of the brane
glued together. The dashed gray lines represent the continuation
of the brane embedding in the absence of the DW and are meant to
indicate how to construct the solutions.
The center diagram shows the locally $AdS$ solution, for which
there is no radiation($p=0$). For positive DW tension, the brane
sweeps more than half of the disk and the remaining opening angle
$\Delta\theta_{opening}\simeq\Delta\eta'$ is less than $\pi$.
Hence, this solution is not asymptotically global $AdS_5$.
The left (right) diagram corresponds to a solution with a positive
(negative) tension DW and asymptotically global $AdS_5$ boundary
condition. Again, the opening angle is less (more) than half of
the original disk. But given that the length of the $S_1$ -- the
boundary of the disk -- can differ from $2\pi$, one can have
asymptotically global $AdS$ if the $S_1$ is larger (smaller) than
$2\pi$.
%
%
%
For this to happen, the Weyl curvature $\mu$ -- and hence the
Casimir Energy $p$ -- must be positive (negative).
%
}
\label{fig:walls_in_KR} %
\end{figure}

So far, we se that solving the $R$ part of the embedding is
(almost) equivalent to finding the warp factor for the quantum
corrected 4D DWs. Let us see now what we obtain when we work out
$\theta(y)$. From \eqref{gauge} and \eqref{friedmanAdS}, one
arrives at
$$
\theta'=\lp(1-{\epsilon\over2}\rp)\,{R\over \ell_5^2\,f(R)}
$$
This means that the angle $\Delta\theta_{brane}$ swept by the
brane from $y\to-\infty$ to $+\infty$ is
\begin{equation}\label{Dthetabrane}
  \Delta\theta_{brane}= \int_{-\infty}^{\infty} \theta' dy = %
  2\lp(1-{\epsilon\over2}\rp)\, \lp[\int_{R_b}^\infty+\,{\rm sign}\sigma
  \int_{R_b}^{R_0}\rp]
{R\over \ell_5^2\,f(R)} {dR \over R'(R)}
\end{equation}
where again $R_b$ denotes the bounce in the warp factor and
$R'(R)$ is found from \eqref{friedmanAdS}. Note that to leading
order in $\epsilon$, this precisely coincides with the conformal
interval of the 4D geometry, Eq. \eqref{Deta*}. Hence, we see that
the angle $\theta$ is an approximate measure of the conformal
coordinate. This gives a geometric justification to the
identification of the length of the $S_1$ as \eqref{general} in
terms of the conformal intervals for the CFT and the CFT$'$ that
we argued for in Section \ref{sec:cft}.

Furthermore, it becomes very clear how to impose the boundary
conditions in terms of $\Delta\theta_{brane}$. As mentioned
earlier, the bulk consists of two copies of the space
\eqref{bulkmetric} cut along the brane trajectory. Picturing these
spaces as disks as in Fig. \ref{fig:hyperSolitons}, the brane is
going to remove a certain wedge-like shape from the disk. Now, if
one is to impose that the 5D metric away from the brane is
asymptotically \emph{global} $AdS_5$, then the total opening angle
left out by the brane should be $\pi$. In equations, what we have
in general is
\begin{equation}\label{DthetaOpening}
\Delta \theta_{opening}=\Delta \theta -\Delta \theta_{brane} ~,
\end{equation} %
where $\Delta \theta$ is given by \eqref{deltaTheta} in order to
avoid a conical singularity in the bulk. The boundary condition
corresponding to asymptotically global $AdS_5$ corresponds to
setting $\Delta \theta_{opening}=\pi$. Since $\Delta\theta$ is a
function only of $\mu$ (that is, the Casimir energy) and
$\Delta\theta_{brane}$ \emph{also} depends on the DW tension
$\sigma$ (as well as on $\ell_4$), this gives a relation between
$\mu$ and $\sigma$, which is going to agree with that of Section
\ref{sec:cft} to leading order in $\epsilon$, of course.

In fact, it is slightly more illuminating to rewrite
\eqref{DthetaOpening} as
$$
\Delta\theta = \Delta\theta_{brane} + \Delta\theta_{opening}~.
$$
This is the counterpart of Eq. \eqref{self2}. We identify
$\Delta\theta(\mu)$ as the strongly coupled version of
$\Delta\theta^{1-loop}(p)$, $\Delta\theta_{brane}$ with the
conformal interval $\Delta\eta$, and $\Delta\theta_{opening}$ with
$\Delta\eta'$. This clearly shows the relation between the choice
of the boundary condition in the $AdS_5$ bulk and the boundary
condition for the CFT$'$.

Hence, the dual of the logic that let us conclude in Section
\ref{sec:cft} that subcritical DWs radiate particles is the
following. In the presence of a DW, the brane sweeps an angle
$\Delta\theta_{brane}$ that is more than half of disk
corresponding to pure $AdS_5$. In order that the remaining opening
angle stays equal to $\pi$, the only option is that the disk is
actually larger than $2\pi$. This demands that the $\mu$ (and
hence the Casimir Energy) is nonzero.


\begin{figure}[tb]
\begin{center}
  \includegraphics[height=9cm]{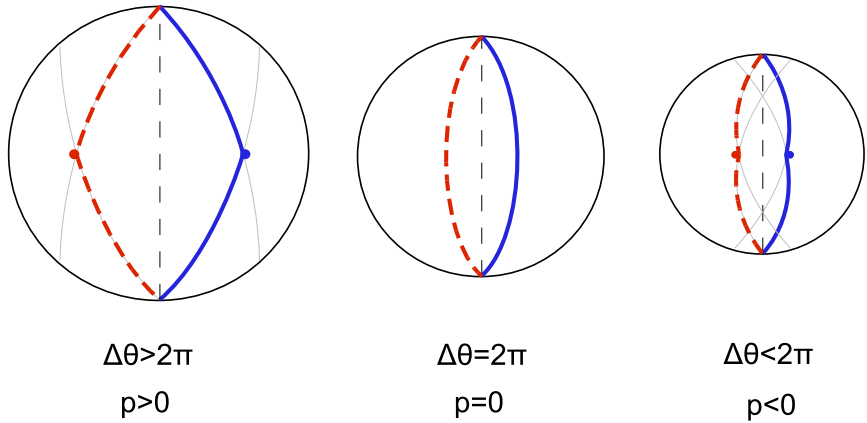}%
\end{center}\vspace{-3cm}
\caption{Schematic view of the localized Domain Walls in the two
brane KR model (corresponding to reflecting boundary conditions).
The straight dashed line indicates the `equator' accross which the
$Z_2$ symmetry is imposed. The solid-blue and dashed-red lines
represent each brane, and the dots the corresponding localized
DWs. The left, right and central diagrams correspond to positive,
negative and zero tension DWs respectively.}
\label{fig:walls_in_KR_refl} %
\end{figure}

In Fig. \ref{fig:walls_in_KR}, we represent schematically how
these localized DW solutions look like. We include for comparison
the case when the bulk is asymptotically global $AdS_5$ and the
case when it is locally $AdS_5$. In these diagrams, we also depict
the embedding of another `fake' brane with the same tension at the
`opposite side' of the bulk would take. One can think that the
CFT$'$ is defined on that brane. For the global $AdS$ boundary
condition, the geometry on the fake brane is conformal to pure
$AdS_4$ (its conformal interval is $\Delta\eta'=\pi$), while for
the locally $AdS_5$
condition, it has a DW$'$ with tension $-\sigma$.\\

Let us now comment on the reflecting boundary conditions. These
are implemented by imposing another $Z_2$ symmetry across the
`equator' of the bulk. In particular this demands the presence of
a second identical brane with an identical DW on it. Hence, in
this case one has
$$
\Delta\theta_{brane}={\Delta\theta\over2}~,
$$
which is the dual of \eqref{reflecting}. The corresponding
solutions are depicted in Fig. \ref{fig:walls_in_KR_refl}.\\

Finally, we should refer to what are the duals of the `Bootstrap'
Domain Wall spacetimes. As explained in Section \ref{sec:qs}, in
the 4D picture when one includes the backreaction from the Casimir
energy, there are DW solutions with with subcritical tension ({\em
i.e.}, with tension smaller that $\sigma_c$ as given by
\eqref{sigmac}) that are planar or even inflating. The curvature
scale of these solutions is well below the cutoff everywhere, so
they can be trusted. Since the KR model is a UV completion of the
4D gravity + CFT system, analogous solutions should be present,
and indeed they are.

The duals of the Bootstrap DW solutions simply are localized DW
solutions like those depicted in Figs. \ref{fig:walls_in_KR} and
\ref{fig:walls_in_KR_refl}, but where one picks the $\kappa=1$ or
$\kappa=0$ slicing of the bulk, for either inflating or planar DWs
respectively.
Since, as already emphasized, the angular coordinate $\theta$ in
the bulk coincides with the conformal coordinate to leading order
in $\epsilon$, the analysis done in Section \ref{sec:qs} already
guarantees that for small $\epsilon$ there exist localized planar
or inflating DW solutions with subcritical tension (one expects a
similar behaviour for inflating DWs). As we have seen in Sec
\ref{sec:qs}, these solutions exist when the DW tension is
comprised between $-\sigma_c$ and an $\epsilon$-dependent maximal
value $\sigma_{planar}^{max}(\epsilon)$, see Fig
\ref{fig:planarQDW}. In the 5D gravity dual,
$\sigma_{planar}^{max}(\epsilon)$ differs from that obtained in
the CFT picture for two reasons: first, because of the large 't
Hooft coupling and second, because the dependence on the
`backreaction parameter' $\epsilon$
only coincides for $\epsilon\ll1$.
In particular, one can imagine that for moderately small values of
$\epsilon$ the duals of the Bootstrap DW solutions `disappear'.
The large 't Hooft coupling can be easily accounted for by taking
the form of $p(\Delta\theta)$ at strong coupling, as has already
been done in Sec \ref{sec:qs} in the planar case. As shown in Fig.
\ref{fig:planarQDW}, the range of DW tensions leading to Bootstrap
solutions is smaller. Hence, in a sense, one can say that strong
coupling effects `remove' some of the solutions where the quantum
effects are important. However, this seems a bit marginal effect.

On the other hand, to see whether these solutions disappear for
larger values of $\epsilon$, one should obtain
$\sigma_{planar}^{max}(\epsilon)$ by solving \eqref{DthetaOpening}
rather than \eqref{self2}. A more detailed study is deferred for
the future, but preliminary results indicate that for modelately
small $\epsilon$ the picture suggested by Fig \ref{fig:planarQDW}
is not significantly changed.


\section{Epilogue: Cosmic Strings in AdS} \label{sec:CS}


Let us finally discuss the gravitational field created by a Cosmic
Strings, that is a relativistic (pure tension) codimension 2
brane, in asymptotically $AdS$ space. This apparently unrelated
issue will turn out to give quite interesting insight. The
following discussion can be done in arbitrary number of dimensions
$D\geq4$, but for the sake of simplicity we shall restrict to the
5D case, in which the Cosmic Strings worldsheet is 2+1
dimensional.

As is well known, it is very easy to obtain the gravitational
effect of a CS in any ambient spacetime in the thin wall
approximation as long as one has a symmetry axis
\cite{vilenkinBook}. The usual prescription is to cut a wedge,
which can simply be done by reducing the range of the angular
polar coordinate. For example, for $AdS$ in Poincar\'e coordinates
\begin{equation}\label{false} {\ell_5^2\over
w^2}\lp[dw^2 -dt^2 + dz^2 + d\rho^2 + \rho^2 d\phi^2 \rp]~,
\end{equation} %
if the range of the angle is $0<\phi<2\pi-\delta$ then this is the
metric for a CS with tension $\sigma = M^3 \delta$. It is also
obvious that locally this metric is isometric to pure $AdS$, and
it only differs from it globally.
It follows that for the solution \eqref{false} the Newtonian
potential vanishes and so the CS exerts no attraction/repulsion on
test particles. Moreover, the Weyl curvature also vanishes
identically everywhere (except at the location of the string, of
course), and so there are no tidal forces either.

However, we shall now argue that \eqref{false} does \emph{not}
represent the actual gravitational field produced by a CS in
$AdS$. The reason is that the boundary conditions for such an
object in $AdS$ are not unique, and as we shall see, those
implicitly assumed to obtain \eqref{false} are not the most
natural ones. Indeed, far away from the CS, the metric
\eqref{false} does not approach global $AdS$. However, it is very
easy to explicitly construct another CS solution that is
asymptotically global $AdS$ and which enjoys the same symmetries.
This is simply given by one of the hyperbolic $AdS$ solitons
\eqref{bulkmetric} (with $\kappa=-1$) with a wedge removed. As we
have seen in the previous Section, once $\mu\neq0$, the length of
the $S_1$ factor $\Delta\theta$ must be given by
\eqref{deltaTheta} in the hyperbolic $AdS$ soliton if one is to
avoid a conical singularity at the center. But that is precisely
what we want for a CS solution. Hence, by choosing $\Delta\theta$
differently from \eqref{deltaTheta}, we will obtain one such CS
solution. For an asymptotically globally $AdS$ CS solution, we
only need to choose $\Delta\theta=2\pi$ (with  $\mu\neq0$).
Equivalently, one can think that we are removing a wedge from a
regular AdS Hyperboloid, in such a way that the length of $S_1$
equal to $2\pi$ and at the same time have a CS at the origin
($R=R_+$). The resulting `witch-hat'\footnote{We thank Lorenzo
Sorbo for suggesting this term to us.} geometry is essentially the
same as the Hyperbolic $AdS$ soliton but with a conical tip at the
center.

Note that the reason why it is possible to find CS solutions that
do not affect the asymptotic structure of the spacetime in $AdS$
is that the Hyperbolic $AdS$ soliton itself can be viewed as a
regular pure gravity cosmic string-like solution, that is a
codimension 2 lump of (Weyl) curvature. The deficit angle produced
by the `material' Cosmic String can thus be compensated by that
produced by the $AdS$ soliton. The Weyl curvature, of course is
not compensated for and this is the effects that remains
at infinity.\\

Now that we know the asymptotically globally $AdS$ CS metrics, let
us describe its properties. With the inclusion of the CS, the
relation between the range of the angular coordinate
$\Delta\theta$, the integration constant $\mu$ and the CS tension
is
$$
{\ell_5 f'(R_+)\over2}\,\Delta\theta=\lp(2\pi-{\sigma\over
M^3}\rp)~.
$$
The metric \eqref{false} corresponds to \eqref{bulkmetric} with
$\mu=0$ (and hence $\Delta\theta=2\pi-\sigma/M^3$), written in
more common coordinates.
For the asymptotically globally $AdS$ metric, instead,
$\Delta\theta=2\pi$ and the Weyl curvature is given in terms of
the CS tension as
\begin{equation}\label{tidalForce}
  {\mu}={\ell_5^2\over32}\lp(8-4 (1-\wh\sigma)^2-(1-\wh\sigma)^4 -(1-\wh\sigma)^3 \sqrt{8+(1-\wh\sigma)^2} \rp)
\end{equation}
where $\wh\sigma\equiv \sigma/(2\pi M^3)$. Note that when the
deficit angle becomes close to $2\pi$ ($\wh\sigma=1$), $\mu$
approaches the extremal value $\mu=4\ell_5^2$, for which the
metric develops an infinitely long throat. This is nothing but
expected, because in this case the geometry close to the string
should be close to a thin cylinder.

The magnitude of the Weyl curvature
$|W^{\mu\nu\rho\sigma}W_{\mu\nu\rho\sigma}|^{1/2}$ is of order
$\mu/R^4$. Hence, it is clear that in this solution the CS
produces long range Weyl curvature and hence tidal forces.
Notice that tidal forces can be produced by wiggling or moving CSs
(as well as by non-relativistic ones). The surprising thing here
is that they are produced in a static, `straight' (maximally
symmetric) configuration. Hence, we should ascribe this as a real
gravitational effect of the strings. The scale of the tidal forces
close to the CS location is of order
$$
{\mu\over R_+^4}\sim{1\over \ell_5^2} {\sigma \over M^3}
$$
for small $\sigma$. Hence, this effect is sensitive to the
curvature scale produced by `the other' sources (in this case, the
cosmological constant), which is why we do not have it for
$\Lambda=0$.\footnote{It is possible to construct solutions with
Weyl curvature (and maximal symmetry) with $\Lambda>0$. However,
it is not clear that these are relevant at all as happens for
$\Lambda<0$, because the asymptotic structure of the resulting
spacetime always differs from de Sitter. It would be interesting
to understand under what circumstances the CS can generate
nontrivial gravitational effects as those presented here. A more
thorough analysis will be left for the future.}

Let us now show that in these solutions the CS generates a
nontrivial Newtonian potential that translates into a repulsive
force on test particles. A quick way to see this is by writing the
metric \eqref{bulkmetric} in the Schwarzschild-like coordinates
\begin{equation}\label{csSchw}
ds_5^2=\ell_5^2 f[R(r)] d\theta^2 + {dr^2\over g(r)} + g(r)
ds_{AdS_3}^2~.
\end{equation}
Taking the $AdS_3$ slices in global coordinates,
$-(1+(x/R_+)^2)dt^2+dx^2/(1+(x/R_+)^2)+x^2d\phi^2$, the Newtonian
potential is simply given by
$$
2\phi_N={g(r)}(1+(x/R_+)^2) -1~.
$$
Comparing \eqref{csSchw} to \eqref{bulkmetric}, one sees that
$g(r)=R^2(r)/R_+^2$ and
\begin{equation}\label{rR}
r(R)=\int_{R_+}^R {R'dR'\over \ell_5 \sqrt{f(R')}}~,
\end{equation}
which can be written explicitly in terms of Elliptic functions.
For our purposes, though, it will suffice to know the form of
$r(R)$ for small and large $r$. At large distances, one finds
$$
r\simeq R + c  +O(R^{-1})
$$
where $c$ is a $\mu$ dependent constant. It is easy to show that
for small $\mu$,
$$
c\simeq{\pi\over4}{\mu\over\ell_5}~.
$$
From this, one immediately finds that
$$
{g(r)} =  
{r^2 -2 c r  \over \ell_5^2}\lp(1+O(\mu/\ell^2)\rp) + \dots~,
$$
where the dots indicate subleading terms in $1/r$. Hence, at
linear level in the CS tension, the Newtonian potential develops a
linear component far away from the CS, corresponding to a constant
gravitational repulsion (for $\sigma>0$) given by
$$
{\pi\over4} { \sigma \over M^3} {1\over \ell_5}~.
$$
It is perhaps surprising to find that a positive tension CS
induces a repulsive force. Intuitively, the reason for this seems
to be that in order to cancel the deficit angle at infinity we
have to start with an $AdS$ soliton with opposite tension, and
this is what really gives the Newtonian potential.

On the other hand, for small $r$, one has $r^2\simeq
{(R^2-R_+^2)/(1-(R_-/R_+)^2)}+\dots$, where $R_\pm$ are the two
roots of $f(R)=0$ and the dots indicate higher powers of
$R^2-R_+^2$. Hence,
$$
{g(r)\over g(0)} =  {R^2 \over R_+^2} = 1+ b^2 \,r^2 +\dots
\qquad{\rm with }\qquad b^2={R_+^2-R_-^2\over R_+^4}~.
$$
Hence, the Newtonian potential is quadratic as usual in $AdS$.
Furthermore, for small $\mu$ one has
$$
b^2\simeq {1\over \ell_5^2}-{\mu^2\over\ell_5^4}+\dots
$$
so, at linear level in the CS tension, the Newtonian potential
close to the CS precisely coincides with that of the usual locally
$AdS$ solution. This is nothing but expected, since in this region
the geometry is approximately conical, and this already accounts
for the gravitational effect of the CS at linear level.\\

In summary, the picture is that for a CS with small tension, close
enough to the CS everything looks `normal', {\em i.e.}, it
generates no Newtonian potential, the Weyl curvature is
essentially irrelevant and all the gravitational effects arise
because the geometry is locally conical. However, after a certain
distance, the Weyl curvature starts to become important, the
notion of the conical-type space is lost and the deficit angle is
replaced by a gravitational repulsion. Naturally, the scale where
the change of regime occurs is given by the curvature radius
associated to the Weyl curvature, which for small $\sigma$ is
$$
{\ell_5\over\sqrt{|\sigma|/M^3}}~.
$$

It is impossible not to notice the quite striking similarity
between this and the screening effect that one would expect in a
massive gravity theory, even though here we have only assumed
ordinary gravity in $AdS$.
Of course, the details of a really massive gravity theory like
$AdS$ plus a CFT with KR boundary conditions are different, but we
already see that even without the CFT there are some elements in
common.
Indeed, in a sense, the deficit angle at infinity is completely
`screened away', and instead, a new kind of behaviour appears at
large distances.
Furthermore, such a change of behaviour only appears because we
are insisting in having a certain kind of boundary condition --
asymptotically global $AdS$.
It is unclear to us whether there is any deep reason why the
gravitational effect of a CS in $AdS$ should display this
massive-like fashion, but it might be that it is a peculiarity
of codimension-2 sources only.\\

So far, we have seen that aside from the ordinary locally $AdS$ CS
solutions, there is another one with the same symmetries that is
asymptotically global $AdS$, with rather different properties.
From a physical point of view, then, we should ask from which
solution should we extract the actual gravitational bahaviour of
the CS in $AdS$, or in other words, what boundary condition should
one impose? Of course this choice is context-dependent, but still
in a generic case one should be able to specify what is the most
natural choice.

For this purpose, it seems quite clear to us that the
asymptotically globally $AdS$ should be preferred.
Generically, whenever one finds several solutions for the same
given source, and one of them shares the same asymptotic structure
as the background, then this seems preferred. In terms of the
linearized theory around that background, presumably this means
that for this solution the metric perturbation is normalizable. In
principle, it may well be (as happens with the CS in
asymptotically flat space) that there is no normalizable solution.
In that case, there is no alternative and and the CS will change
the global structure at infinity. But in $AdS$ this does not need
to be so: the persistence of the wedge at infinity can be traded
by the presence of a long range Wely curvature.
From a physical point of view, one can always consider situations
where the cosmic string is produced dynamically by some mechanism
localized in some finite region. If initially the space is
asymptotically global $AdS$, then once the CS is created one
should still have the same asymptotics.\\

%

Finally, let us point out that the curvature scale on the string
is  $-1/R_+^2$ with $R_+$ given in \eqref{Rplus}, so this also
depends on the string tension. Hence, one obtains a nontrivial
`Friedmann equation' (understood as the relation between the
curvature scale and the tension or energy density on the defect;
this is more relevant for the generalization of the present
example to 6D, in which case the CS would be a 3+1 brane). In
fact, the Friedmann equation would be quite peculiar, because the
larger $\sigma$ is, the more negative the curvature scale becomes.
This is not so strange of course because one does not have lower
dimensional gravity localized on the defect. However, this is
illustrative as an extreme case where even with a noncompact bulk
there is no self-tuning mechanism at work (once the asymptotically
globally $AdS$ boundary condition is enforced).

\section{Conclusions}
\label{sec:concl}

\begin{itemize}

    \item
    A relativistic Domain Wall (DW) is a physical implementation of
    a uniformly accelerating mirror. In $AdS_4$, whenever the DW
    acceleration exceeds the curvature scale $1/\ell_4$,
    there is no particle production into conformal fields in
    the maximally symmetric configuration (both at weak and strong coupling).
    However, subcritical DWs (with acceleration $<1/\ell_4$) generate a
    nontrivial CFT particle
    production that can be in equilibrium with the DW and the
    $AdS_4$ boundary.

    \item The amount of CFT radiation produced by subcritical DWs in $AdS_4$ is not
    dramatically sensitive to wether the CFT is weakly or
    strongly coupled (see Fig. \ref{fig:T_p_ratios}). With Karch-Randall (KR) boundary conditions, the energy
    density in the radiation at weak and at strong coupling agree within a few
    \%, and for reflecting boundary conditions the maximum
    discrepancy is a factor $3/4$.

    \item The Karch-Randall type boundary conditions really represent a family of
    conditions. They encode the choice of boundary conditions for the CFT
    in the choice of vacuum for the `adjacent' CFT$'$.
    The only natural choice is that the CFT$'$ is in the \emph{ground
    state} in a space conformal to pure $AdS_4$, which corresponds in the 5D dual to
    choosing an
    asymptotically \emph{globally} $AdS_5$ bulk.
    Departing from this condition can lead
    to considerably different amount of radiation (both at weak and strong coupling),
    but this corresponds to a CFT in a quite exotic state.

    \item Once the boundary conditions between the two sides are
    appropriately matched, the DW solutions with the backreaction
    from the CFT included
    agree with the solutions for DWs localized on the
    brane in the Karch-Randall model to leading order in the
    backreaction. This confirms the holographic interpretation of
    the KR model as a CFT coupled to 4D (semiclassical) gravity in
    $AdS_4$.

    \item We found a new type of solutions (the `Bootstrap' DW spacetimes) which exist
    thanks to the backreaction from $\tmn^{CFT}$ (for $\Lambda_4<0$).
    These solutions have no classical
    analog. Yet, they have a 5D gravity dual, further confirming
    the cutoff AdS/CFT correspondence.

    \item Our results also comply with the massive gravity interpretation
    of the Karch-Randall setup.
    The backreaction of the produced radiation
    leads to a screening of the DW tension, which depends on the boundary conditions.
    For KR boundary conditions there is more
    screening than for reflecting boundary conditions, as expected since
    the KR choice leads to a nonzero graviton mass.

    \item The phenomenon dual to the particle creation by
    subcritical DWs in AdS is that pure-tension codimension 2 branes in AdS
    are repulsive and
    produce long range Weyl curvature, {\em i.e.}, tidal forces.
    This effect is quite interesting by itself, since this is
    rather
    different from the usual behaviour in flat space.
    It would be interesting to see whether it is peculiar to
    $AdS$ or whether it also happens more generically whenever a
    codimension 2 brane is placed in an already curved space.

\end{itemize}

\section*{Acknowledgements}

I thank JJ Blanco-Pillado, G Dvali, G Gabadadze, J Garriga, L
Grisa, A Iglesias, M Kleban, M Porrati, M Redi and T Tanaka for
useful discussions and R Emparan for very valuable comments on a
previous version of the paper. I also thank the organizers and
participants of the ``Quantum Black Holes, Braneworlds and
Holography" workshop (Valencia, May 12-16 2008) for their
feedback.
This work has been partially supported by DURSI under grant 2005
BP-A 10131 and the David and Lucile Packard Foundation Fellowship
for Science and Engineering

\bibliographystyle{utphys}
\bibliography{AccMirr}

\end{document}